\documentclass
[aps,prc,twocolumn,showpacs,showkeys,amsmath,floatfix,superscriptaddress]{revtex4}
\usepackage[dvips]{color}
\usepackage{color,graphicx,xcolor}
\usepackage{subfigure}
\usepackage[utf8]{inputenc}
\usepackage{mathptmx}                % math+times fonts
\usepackage{dcolumn}                 % Align table columns on decimal point
\usepackage{bm}                      % bold math

\begin{document}

%opening
\title{EoS constraints from nuclear masses and radii in a meta-modeling approach}

\author{D. Chatterjee$^1$\thanks{dchatterjee@lpccaen.in2p3.fr}, F. Gulminelli\thanks{gulminelli@lpccaen.in2p3.fr}}
\affiliation{LPC, UMR6534, ENSICAEN, F-14050 Caen, France}
\author{Ad. R. Raduta\thanks{araduta@nipne.ro}}
\affiliation{IFIN-HH, Bucharest-Magurele, POB-MG6, Romania}
\author{J. Margueron\thanks{j.margueron@ipnl.in2p3.fr}}
\affiliation{Institute for Nuclear Theory, University of Washington, Seattle, Washington 98195, USA}
\affiliation{Institut de Physique Nucl\'eaire de Lyon, CNRS/IN2P3, Universit\'e de Lyon, Universit\'e Claude Bernard Lyon 1, F-69622 Villeurbanne Cedex, France}

\date{\today}

\begin{abstract}
The question of correlations among empirical equation of state (EoS) parameters constrained by nuclear observables is addressed in a fully empirical meta-modeling approach.  {A recently proposed meta-modeling for the nuclear EoS in nuclear matter}, is augmented with a single finite size term to produce a minimal unified EoS functional able to describe the smooth part of the nuclear ground state properties. This meta-model can  reproduce the predictions of a large variety of models, and interpolate continuously between them. The parameter space is sampled and filtered through the constraint of nuclear mass reproduction. We show that this simple  analytical meta-modeling has a predictive power on masses, radii, and skins comparable to full HF or ETF calculations with realistic energy functionals. 
The covariance analysis on the posterior distribution shows that no physical correlation is present between the different EoS parameters. 
Concerning nuclear observables, a strong correlation between the slope of the symmetry energy and the neutron skin is observed, in agreement with previous studies.
\end{abstract}

\maketitle

\section{Introduction}

The nuclear Equation of State (EoS) is one of the fundamental entities that governs the behavior of nuclear systems, from terrestrial nuclei to neutron stars~\cite{oertel16}. However, astrophysical observations alone cannot provide enough information to constrain the behavior of asymmetric nuclear matter in the different density and isospin asymmetry domains, and the best knowledge of the EoS still comes from low energy nuclear physics experiments, i.e., nuclear ground state properties such as masses and radii of nuclei or the neutron skin, among others~\cite{epja50,Lattimer16}. 

For this reason, extracting reliable confidence intervals for the EoS empirical parameters from the laboratory data has become a crucial issue in compact star modelling. Thanks to the new developments in density functional theory (DFT), ab-initio modelling, and to the data on asymmetric nuclei collected from the new rare ion beam facilities worldwide, enormous progress was achieved in the past few years~\cite{tsang_rep,lattimer_lim,lattimer_steiner,dutra14,steiner_bayes,fortin16}.

Despite the huge advancement, some methodological issues are still open. We wish to contribute to two such issues in this paper.

First, the EoS parameters are not direct observables, and experimental data are typically fitted by theoretical models based on phenomenological functionals which assume specific functional forms for the effective  Hamiltonian or Lagrangian terms.
As a consequence, the different EoS parameters extracted from the data fit exhibit strong correlations~\cite{Kortelainen12,Nazarewicz14,Danielewicz14,McDonnel15}. Some of these correlations might reflect physical properties of the EoS. This would be for instance the case if experimental constraints concern two well distinguished different density domains: imposing that the functional passes through two different points  implies some obvious correlations among the successive derivatives of the energy functional at a given density point. It is however highly probable that many of such correlations might be spuriously induced by the insufficient exploration of the complex parameter hyper-surface, and the specific functional form assumed for the EoS, which as long as phenomenological models are used has always a certain degree of arbitrariness. The presence of spurious effects is clearly indicated by the fact that the correlation among empirical parameters are modified by changing the functional family, from Skyrme to Gogny or RMF or RHF~\cite{Nazarewicz14,Ducoin2011,KhanMargueron,Casali1,Mondal}.  A simple example is given by non-relativistic functionals: it is easy to show analytically that the phenomenological density dependent term of Skyrme interactions induces artificial correlations among the different isoscalar as well as isovector EoS terms~\cite{KhanMargueron}. 
These functional correlations are then propagated to the analysis of the correlations among the nuclear observables and the empirical parameters, affecting the estimation of the confidence intervals and therefore the predictions of neutron star observables. 

A second problem comes from the fact that the energy functionals used to describe nuclear properties contain many more parameters than the ones entering the EoS of infinite nuclear matter, such as (at least) effective masses, surface, spin-orbit and tensor terms. The proliferation of extra parameters, though essential to pin down the complexity of nuclear structure, makes it hard to sort out the specific effect of EoS parameters on a given nuclear observable~\cite{Nazarewicz14}. To give an example, this problem is  shown by different recent works analyzing the EoS dependence of the neutron skin in $^{208}$Pb. Based on a droplet model analysis, the Barcelona group~\cite{warda09} has shown that the skin is not directly correlated to the EoS parameter $L_{sym}$ (slope of the symmetry energy at saturation), 
but rather to the ratio $L_{sym}/Q$ where $Q$ represents the so called surface stiffness parameter, which is related to the differential surface tension between protons and neutrons. Consequently, one can expect that the EoS independent surface contribution to the skin should be sorted out  in order to recover the correct correlation with the EoS. However, when using more sophisticated EoS models it  was shown~\cite{Mondal,centelles10,reinhard16} that the surface contribution is remarkably constant in the different relativistic and non-relativistic models, and therefore the correlation of the skin with $L_{sym}$ is preserved. It was conjectured~\cite{centelles10} that this constant behavior might come from the constraint on nuclear mass reproduction applied to the DFT functionals, but the argument neglects the fact that the mass itself is known to provide constraints on the EoS~\cite{tsang_rep,kortelainen10,Kortelainen12} besides the constraints on the finite size parameters~\cite{Jodon} .
 
To progress on these issues, we will consider in this paper a meta-modeling for the nuclear equation of state inspired  by a Taylor expansion of the energy density around saturation~\cite{Casali1}, augmented by an isoscalar surface term, which can be taken as an effective representation of finite size effects. 
 {With this extra term, the concept of meta-modeling introduced in nuclear matter in refs.~\cite{Casali1,Casali2} is here extended to finite nuclei.}
We show that this single parameter is sufficient to pin down the full isoscalar and isovector behavior of the smooth part of the nuclear binding energy, and the predictive power of the model is not significantly improved if further parameters are added. 

The use of a Taylor expansion allows a full and even exploration of the multidimensional EoS parameter  space, without any a-priori correlation in the functional form~\cite{Casali2}. The addition of a single finite size term, subject to the constraint of correctly reproducing experimental nuclear mass, allows to single out the interplay between bulk and surface parameters. 

With this meta-modeling, we will then sample the multi-dimensional parameter space with a prior flat distribution, and filter the parameter set through the constraint of least-square mass reproduction. This allows to explore the possible physical correlations among empirical parameters and carry out a systematic investigation of radii and neutron skins
in nuclei. We will show that the slope of symmetry energy strongly correlates with the neutron skin, and its measurement is virtually unaffected by the uncertainty in any of the other empirical quantities, thus highlighting the importance of neutron skin in the determination of the nuclear EoS~\cite{Brown,Typel,Piekarewicz}.

\section{Formalism} \label{sec:hom}

{In this section, we shortly review the formalism of meta-modeling for uniform nuclear matter. Then we detail its implementation in finite nuclei within an analytical Extended Thomas-Fermi mass model (ETF) where the parameters of the density profile are directly related to the empirical parameters of nuclear matter. 
A complete ETF solution of the problem where the parameters  of the density profile are fully varied is also introduced.}

\subsection{Meta-modeling for homogeneous matter}\label{sec:emp_eos}

To describe homogeneous nuclear matter, we consider a meta-modeling where the  EoS  
coefficients are directly related to the state-of-the-art knowledge of nuclear matter based
on data from nuclear experiments. The energy per particle in asymmetric nuclear matter can be separated into isoscalar and isovector channels, as
\begin{equation}
e(n_0,\delta) = e_{IS}(n_0) + \delta^2 e_{IV}(n_0)~.
\label{eq:expansion}
\end{equation}
Here, $\delta=n_1/n_0$ is the asymmetry of bulk nuclear matter, the density $n_0=n_n+n_p$ (resp. $n_1=n_n-n_p$)
being the sum (resp. difference) of proton and neutron densities $n_p$ and $n_n$.
In principle the expansion in asymmetry parameter $\delta$ in Eq.~(\ref{eq:expansion}) can go up to any even power of $\delta$, due to isospin symmetry.
In practice stopping at order $\delta^2$ is usually enough~\cite{Casali1} for the potential part.
Non-quadraticities in the kinetic part are treated explicitly, see below.
The empirical parameters appear as the coefficients of the series expansion around saturation density $n_{sat}$
in terms of a dimensionless parameter $x = (n_0 - n_{sat})/(3 n_{sat}) $, i.e.,
\begin{eqnarray}
e_{IS} &=& E_{sat} + \frac{1}{2} K_{sat} x^2 + O(x^3)   \label{eq:e_is} \\
e_{IV} &=& E_{sym} + L_{sym} x + \frac{1}{2} K_{sym} x^2 + O(x^3). \label{eq:e_iv}
\end{eqnarray}
The isoscalar channel is written in terms of the energy per particle at saturation $E_{sat}$, and the isoscalar incompressibility $K_{sat}$. 
{There is no linear term in $x$ since the pressure is zero at saturation density.}
The isovector channel is defined in terms of the symmetry energy $E_{sym}$ and its first two derivatives $L_{sym}$, $K_{sym}$.  
In principle, there is an infinite number of terms in the series expansion. 
{To reproduce well the equation of state and its derivatives up to $4n_{sat}$, it is enough to stop at order 4 in $x$~\cite{Casali1}.
For our purpose, since we are only interested in finite nuclei, around and below saturation density, it is sufficient to stop
at order 2 in $x$~\cite{Casali1}.}
Therefore in this study, which is uniquely centered on nuclear ground state properties, we do not consider higher order terms.\\

To achieve a faster convergence at low density and large isospin asymmetries, the density dependence of the kinetic energy term is separated from that of the potential term:
\begin{equation}
e (x,\delta) = e_{kin} (x,\delta) + e_{pot} (x,\delta)~.
\label{eq:e_hnm}
\end{equation}

The kinetic energy term can be written as:
\begin{equation}
 e_{kin} = t_{sat}^{FG} (1+3 x)^{2/3} \frac{1}{2} \left[ (1+\delta)^{5/3} \frac{m}{m_n^*} + (1-\delta)^{5/3}\frac{m}{m_p^*}  \right]~,
\end{equation}
where  the constant $t_{sat}^{FG}$ is given by:
\begin{equation}
 t_{sat}^{FG} = \frac{3}{5}\frac{\hbar^2}{2m} \left( \frac{3 \pi^2}{2} \right)^{2/3} n_{sat}^{2/3} ~,
\end{equation}
and $m_n^*$ and $m_p^*$ are the in-medium masses of protons and neutrons, $m$ being the bare nucleon mass. 
{The in-medium effective mass can also be expanded
in terms of the density parameter $x$ as~\cite{Casali1}:
\begin{equation}
\frac{m}{m^*_q} = 1+ \left( \kappa_{sat} + \tau_3 \kappa_{sym} \delta \right) \frac{n_0}{n_{sat}},
%\sum_{\alpha=0}^1 m_{\alpha}^q (\delta) \frac{x^{\alpha}}{\alpha!}~.
\label{eq:effm_expn}
\end{equation}
where $\tau_3=1$ for neutrons and $-1$ for protons.

This expansion adds two extra empirical parameters to our parameter set, namely:
\begin{eqnarray}
\kappa_{sat} &=& m/m_{sat} - 1 \nonumber \\
\kappa_{sym} &=& \frac{1}{2} [m/m_n^*(x=0,\delta=1)-m/m_p^*(x=0,\delta=1)]. 
\label{eq:effm_def}
\end{eqnarray}
 Here, $\kappa_{sat}$ is related to the isoscalar effective mass and $\kappa_{sym}$ is half the difference of the inverse of the effective masses in
 neutron matter.
 More commonly, the isovector dependence of the effective mass is described in terms of the isospin splitting of the nucleon masses in neutron matter,
\begin{eqnarray}
\frac{\Delta m^*}{m} &=& \frac{1}{2} [m_n^*(x=0,\delta=1)-m_p^*(x=0,\delta=1)]. \nonumber \\
&=& \frac{\kappa_{sym}}{(\kappa_{sym})^2-(1+\kappa_{sat})^2}.
\label{eq:effm_def}
\end{eqnarray}
The value of $\Delta m^*/m$ is not very well constrained from experimental data~\cite{Lesinski2006}. 
Theoretical predictions based on Bruckner-Hartree-Fock formalism prefer a small value of the order of 0.1~\cite{Lesinski2006}.
In this work, we fix $\kappa_{sym}=0$ since this parameter has a very weak effect for masses of finite nuclei.
}

Similarly, we may write the potential part of the energy per particle as a series expansion separated into isoscalar and isovector contributions $a_{\alpha 0}$ and $a_{\alpha 2}$, up to second order in the parameter $x$ as follows:
\begin{equation}
e_{pot} = \sum_{\alpha=0}^2 (a_{\alpha 0} + a_{\alpha 2} \delta^2) \frac{x^{\alpha}}{\alpha!} u_{\alpha} (x) \,,
\label{eq:epot}
\end{equation}
{This expression for the potential term corresponds to the meta-modeling ELFc of Ref.~\cite{Casali1}.
The function $u$ is defined as $u_{\alpha} (x) = 1 +27x^3 e^{-b(3x+1)}$ such that $e_{pot}$ satisfies the following limit:
$e_{pot}\rightarrow 0$ for $n_0\rightarrow 0$.
The parameter $b$ is set to $b=10 \ln 2$ such that it has a negligible contribution above saturation density, see Ref.~\cite{Casali1} for more details.}

Then comparing with Eqs.~(\ref{eq:e_is}) and (\ref{eq:e_iv}), the isoscalar and isovector coefficients in the expansion can be written in terms of the 
empirical parameters as~\cite{Casali1},
\begin{eqnarray}
a_{00} &=& E_{sat} - t_{sat}^{FG} ( 1 + \kappa_{sat}) \\
a_{10} &=& - t_{sat}^{FG} ( 2 + 5 \kappa_{sat}) \\
a_{20} &=& K_{sat} - 2 t_{sat}^{FG} ( -1 + 5 \kappa_{sat} ) \\
a_{02} &=& E_{sym} - \frac{5}{9} t_{sat}^{FG} ( 1 + \kappa_{sat}) \\
a_{12} &=& L_{sym} - \frac{5}{9} t_{sat}^{FG} ( 2 + 5\kappa_{sat}) \\
a_{22} &=& K_{sym} - \frac{10}{9} t_{sat}^{FG} ( -1 + 5\kappa_{sat}) 
\end{eqnarray}

\subsection{Analytical mass model}\label{sec:mass_model}

The mass of a finite nucleus is obtained  performing an analytical integration of the local energy functional folded with a parametrized density profile, in the ETF approximation. 
The energy functional is given by the meta-modeling presented in Section~\ref{sec:emp_eos}, {complemented with a gradient term} to account for finite size effects. 
The resulting meta-modeling will be called 
%"empirical EOS" 
 {"meta-functional"}
in the following.

The theoretical method is described in detail in Refs.~\cite{Francois1,Francois2} in the case of a Skyrme functional.
Here we only give the main results and the differences of the present study with respect to \cite{Francois1,Francois2}, arising from the use of the meta-functional,
instead of the Skyrme functional.
Given a parametrized form for the proton and neutron density profiles $n_q(r)$ ($q =n,p$), the nuclear part of the energy of a spherical nucleus can be determined by integrating in space the local energy functional in the ETF approximation: 
 
\begin{equation}
 E_{nuc} = 4\pi \int_0^\infty dr r^2 {\cal{H}}_{ETF} [n_p (r), n_n(r)]~.
\label{integral}
\end{equation}

The ETF functional at the second order in $\hbar$ is given by:
\begin{eqnarray}
\mathcal{H}_{ETF}[n_n,n_p] =
 e(n_n,n_p)  n_0
+ \sum_{q=n,p} \frac{\hbar^2}{2m_q^*}  \tau_{2q} 
+  C_{fin} 
 \left( \boldsymbol{\nabla}n_0 \right)^2. \nonumber \\
\label{eq_sym_density_energy_Skyrme}
\end{eqnarray}

Here, the energy per particle of uniform nuclear matter $e$ and the effective masses $m_q^*$ are given by Eqs.~(\ref{eq:e_hnm}) and (\ref{eq:effm_expn}). 
The second order local and non-local corrections $\tau_{2q}=\tau_{2q}^l + \tau_{2q}^{nl}$ are given by the $\hbar$ expansion as:
\begin{eqnarray}
\tau_{2q}^l    & =& \frac{1}{36} \frac{ \left( \boldsymbol{\nabla}n_q \right)^2 }{n_q} + \frac{1}{3} \Delta n_q  \label{tau_loc}\\
\tau_{2q}^{nl} & =&
\frac{1}{6} \frac{ \boldsymbol{\nabla}n_q \boldsymbol{\nabla} f_q}{f_q}
+\frac{1}{6}  n_0  \frac{\Delta f_q}{f_q}
- \frac{1}{12}  n_q  \left(  \frac{\boldsymbol{\nabla} f_q}{f_q} \right) ^2, \label{tau_nloc}
\end{eqnarray}
with $f_q=m/m_q^*$. 
Finally, the $C_{fin}$ term is an extra isoscalar surface term which does not play any role in infinite matter, but has to be added as an extra parameter to our empirical parameter set, when we compute finite nuclei observables \cite{bulgac15}.

The integral in Eq.~(\ref{integral}) can be analytically performed if the density profiles are taken as Fermi functions,
\begin{equation}
 n_q(r) =  n_{bulk,q} F_q(r), \;\; F_q(r) = (1+e^{(r-R_q)/a_q})^{-1}.\label{eq:fermi}
\end{equation}
The  parameters $n_{bulk,q}$ can be obtained from the infinite matter limit of the Euler-Lagrange variational equations \cite{krivine} as the proton and neutron saturation densities \cite{Panagiota}:
$n_{bulk,q} = n_{bulk} (\delta) (1 \pm \delta)/2$, and $R_q$ are univocally fixed by particle number conservation. 
In the above expression, the saturation density for asymmetric matter depends on the asymmetry $\delta$ and can be written as a function of the 
empirical parameters as~\cite{Panagiota}
\begin{equation}
n_{bulk} (\delta) =  n_{sat} \left( 1 - \frac{3 L_{sym}\delta^2}{K_{sat} + K_{sym} \delta^2} \right)~.  \label{rho0}
\end{equation}\\

The diffuseness of the density profiles ($a_n$ and $a_p$) must be variationally determined by energy minimization, as it is customary in ETF calculations.  
{In finite nuclei, there is a difference between the asymmetry parameter $\delta$ and the global asymmetry of the nucleus $I=(N-Z)/A$.
By comparing liquid-drop models based on $\delta$ or on $I$, the best reproduction of self-consistent Hartree-Fock calculations is obtained using 
$\delta$~\cite{Panagiota,Francois3},
which is defined in the framework of the droplet model as~\cite{meyers}:
\begin{equation}
\delta = \frac{I + \frac{3 a_c Z^2}{8 Q A^{5/3}} }{1 + \frac{9 E_{sym}}{4 Q A^{1/3}} }~,
\end{equation}
where $a_c=3e^2/(20\pi\epsilon_0r_{bulk}(\delta))$ is the Coulomb parameter, 
and $r_{bulk}(\delta)= \left(\frac{4}{3}\pi n_{bulk}(\delta)\right)^{-1/3}$ the mean radius per nucleon.
A detailed discussion can be found in Ref.~\cite{Francois3}.

With these definitions, the mass model can be computed if we specify a single extra parameter ($C_{fin}$) in addition to the  {empirical parameter} set $\{P_\alpha\}\equiv \{n_{sat},E_{sat},\kappa_s,K_{sat},E_{sym},L_{sym},K_{sym}\}$.

The total nucleus energy is obtained by direct integration of Eq.~(\ref{integral}) and can be decomposed into a bulk and a surface contribution:
\begin{equation}
E_{nuc}(A,\delta) = E_b + E_s. \label{eq:mass}
\end{equation}
The bulk energy is the equilibrium energy of homogeneous nuclear matter at isospin $\delta$
\begin{equation}
 E_b(A,\delta) = e(n_{bulk}(\delta),\delta) A~,
\end{equation}
where  $e$ is defined in Eq.~(\ref{eq:e_hnm}).
The surface energy $E_s$ contains contributions from the gradient terms in the energy functional. 
It can be further decomposed into an isoscalar and an isovector part,
\begin{equation}
E_s(A,\delta)=E_s^{IS}+E_s^{IV}\delta^2. \label{eq:es}
\end{equation}
  Thanks to the Fermi profile ansatz, the integration of the isoscalar part is fully analytical, giving \cite{Francois1,Francois2}:
\begin{eqnarray}
E_s^{IS} &=& \left ( {\cal{C}}_{surf}^L(\{P_\alpha\})+ {\cal{C}}_{surf}^{NL}(\{P_\alpha\},C_{fin})\right ) \frac{a}{r_{bulk}} A^{2/3}  \nonumber \\
&+&  \left ( {\cal{C}}_{curv}^L(\{P_\alpha\})+ {\cal{C}}_{curv}^{NL}(\{P_\alpha\}, C_{fin})\right )\left[ \frac{a}{r_{bulk}} \right]^2 A^{1/3} \nonumber \\
&+&  \left ( {\cal{C}}_{ind}^L(\{P_\alpha\})+ {\cal{C}}_{ind}^{NL}(\{P_\alpha\},C_{fin})\right ) \left[ \frac{a}{r_{bulk}} \right]^3 ~,
\label{eq:e_s^l}
\end{eqnarray}
where 
 the explicit expressions for the
${\cal{C}}_{surf/curv/ind}^{L/NL}$ coefficients are given in Refs.~\cite{Francois1,Francois2}.

We can see that the surface energy is constituted of a surface, a curvature,  and a constant term \cite{meyers}. 
In turn, these terms can be separated in a local and a non-local part. The local terms depend only on the bulk density $n_{bulk}$ and on the empirical EoS parameters $\{P_\alpha\}$ (we
 recall $n_{bulk}$ depends only on the isospin parameter $\delta$ and on the empirical $\{P_\alpha\}$ set, see Eq.~(\ref{rho0})). The non-local part also depends on the gradient terms and thus on the finite size parameter ($C_{fin}$, in the present application). 
It is also interesting to remark that the isospin dependence of the surface energy is more complex than in the usual parabolic approximation, and the expression (\ref{eq:es}) effectively contains higher orders in $\delta$ because of the $\delta$ dependence of the diffuseness $a_q$, saturation radius $r_{bulk}$, and of the saturation density $n_{bulk}$. 
\\
The isovector surface part can be evaluated in the gaussian approximation as \cite{Francois2}:
\begin{equation}
E_s^{IV} = E_{surf}^{IV} A^{2/3} + E_{ind}^{IV} \,, \label{eq:e_s^nl}
\end{equation}
where again the coefficients only depend on the EoS parameters, $E_{surf}(\{P_\alpha\}, n_{bulk}(\delta))$, $E_{ind}(\{P_\alpha\}, n_{bulk}(\delta))$.
It was shown in Ref.~\cite{Francois2} that this approximation to the surface symmetry energy gives a good reproduction of a full ETF calculation
using the same density profiles, for isospin asymmetries up to $\delta\approx 0.3$ or $I\approx 0.4$, which are very close to the neutron drip-line.

The last parameters of the model are the diffuseness of the density profiles $a_q$. 
These parameters  can be determined by mimimizing the energy, 
i.e. $ \frac{\partial E}{\partial a_q} = 0$. In the simplifying approximation $a_n=a_p=a$ this minimization can be analytically performed and gives \cite{Francois2}

\begin{eqnarray}
a^2 (A,\delta) &=& \frac{{\cal{C}}^{NL}_{surf} (\delta)}{{\cal{C}}^L_{surf} (\delta) }
+ \Delta R_{HS} (A,\delta)\times \label{eq:diffuseness}  \\
&\times& \sqrt{\frac{\pi}{(1-\frac{K_{1/2}}{18 J_{1/2}})}} \frac{ n_{sat}}{ n_{bulk}(\delta)} 
\frac{3 J_{1/2}}{{\cal{C}}^L_{surf}(\delta)} 
\sqrt{\frac{{\cal{C}}^{NL}_{surf}(0)}{{\cal{C}}^L_{surf}(0)} }
 (\delta-\delta^2)~. \nonumber 
\end{eqnarray}

In this expression, the coefficients $J_{1/2}$ and $K_{1/2}$ represent the value of the symmetry energy and its curvature at one half of the saturation density, 
$J_{1/2}=2e_{IV}( n_{sat}/2)$,
$K_{1/2}=18(\frac{ n_{sat}}{2})^2\partial^2 e_{IV}/\partial  n^2 |_{ n_{sat}/2}$, and 
\begin{equation}
\Delta R_{HS}=\left (\frac{3}{4\pi}\right )^{1/3} 
\left [\left (\frac{A}{ n_{bulk}(\delta)}\right )^{1/3}-\left (\frac{Z}{ n_{bulk, p}(\delta)}\right )^{1/3}\right ] \label{eq:rhs}
\end{equation}
 is the difference between the hard sphere radius $R_{HS}=r_{bulk}(\delta)A^{1/3}$ and the proton radius $R_{HS,p}=r_{bulk, p}(\delta)Z^{1/3}$  in the hard sphere limit.

{One should observe that the approximation $a_n=a_p$ employed to obtain Eq.~(\ref{eq:diffuseness}) is not verified in complete variational ETF or HF calculations of asymmetric nuclei, see for instance Refs.~\cite{centelles13,danielewicz17}.
It was indeed suggested that a substantial fraction of the neutron skin is induced by the difference between $a_n$ and $a_p$~\cite{warda09,centelles10,centelles13}.
This effect corresponds to the non-bulk contribution to the neutron skin.
In our model, we can however see from Eq.~(\ref{eq:diffuseness}) that, due to the complex isospin dependence of the ETF functional, the diffuseness $a$ explicitly contains isovector non-bulk contributions generated by the non-local terms.
We will see in Section \ref{sec:radii} that  even within the approximation $a_n=a_p$ the skin in our model acquires a finite surface contribution.
While not being explicitly in contradiction with the arguments presented in Refs.~\cite{warda09,centelles10,centelles13}, our meta-modeling shows that the richness
of the mean-field induces sometimes more features than a-priori expected.}

To compare with experimental binding energy, we also add a Coulomb contribution to the total energy from equations (\ref{eq:e_s^l}) and (\ref{eq:e_s^nl}), which becomes 
\begin{equation}
E_{tot}(A,Z)=E_b(A,\delta)+E_s(A,\delta)+ a_c\frac{Z^2}{A^{1/3}}\label{eq:binding} \, .
\end{equation}
Within the same ETF approximation, we can also analytically calculate the root-mean-square (rms) radii of protons $\sqrt{ \langle r_p^2 \rangle}$ and neutrons $\sqrt{ \langle r_n^2 \rangle}$  as~\cite{centelles10,Panagiota}:
\begin{equation}
\langle r_q^2 \rangle= \frac{3}{5} R_{HS,q}^2
 \left ( 1+ \frac{5\pi^2 a^2}{6 R_{HS,q}^2} \right )^2, \label{eq:radius}
\end{equation}
where the diffuseness $a$ is given by Eq.~(\ref{eq:diffuseness}).
We can see from Eqs.~(\ref{eq:radius}) and (\ref{eq:diffuseness}) that
the radii are  explicitly correlated to $ n_{sat}$ through $r_{bulk, q}$ defining $R_{HS, q}$, but all the other isovector and isoscalar empirical parameters 
including the finite size $C_{fin}$ additionally enter in the radius definition.
{As discussed before, the radii are thus related to all empirical and surface parameters in a complex way.}

The neutron skin is given by the difference in the rms radii of protons and neutrons, i.e.,
$\Delta R_{np} = \sqrt{ \langle r_n^2 \rangle} - \sqrt{ \langle r_p^2 \rangle}$. 
To compare with the observations, one must calculate the charge radius which is related to
the proton radius, using the relation:
$$ \langle r^2_{ch} \rangle^{1/2} = \left[ \langle r^2_p \rangle + S_p^2 \right]^{1/2},$$
where the correction $S_p$ = 0.8 fm comes from the internal charge distribution of the proton \cite{Buchinger,Patyk}. 

With this fully analytical ETF mass model, 
masses and radii can be evaluated for any arbitrary $\{ P_\alpha\}$ set of empirical parameters, 
provided an estimation of the extra finite size parameter $C_{fin}$ is given. The determination of this parameter is detailed in Section \ref{sec:reference}.

\subsection{Full ETF  {mass model}}
\label{sec:full_ETF}

In order to quantify the possible limitations due to the different approximations employed in the analytical mass model, we also present in this work
results of a full variational ETF approach where the binding energy is calculated from the following numerical integration:
\begin{equation}
E_{tot}(A,\delta)=4\pi \int_0^\infty dr r^2 \left ({\cal{H}}_{ETF} [n_p, n_n]+{\cal{H}}_{coul} [n_p]\right ).
\end{equation}
In the full ETF approach, the density profiles are still given by Eq.~(\ref{eq:fermi}), but the parameters $n_{bulk, q}$ and $a_q$ are treated as four 
independent variational variables.
In addition, the minimization with respect to these variables is performed including the direct and exchange Coulomb terms at the level of the energy density 
defined as~\cite{onsi08}:
\begin{eqnarray}
 {\cal{H}}_{coul} [n_p]&=&2\pi e^2 n_p(r)   \int_0^r dr' n_p(r')\left (\frac{r'^2}{r}-r'\right ) \nonumber \\
&-&\frac{3e^2}{4} \left (\frac{3}{\pi}\right)^{1/3}n_p^{4/3}(r) \,,
\end{eqnarray}
where the Slater approximation has been employed to estimate the exchange Coulomb energy density.

\section{Fixing a set of reference parameters} \label{sec:reference}

In the following, we vary the value of the EoS empirical parameters 
$\{P_\alpha\}\equiv \{ n_{sat}$, $E_{sat}$, $\kappa_{sat}$, $K_{sat}$, $E_{sym}$, $L_{sym}$, $K_{sym}\}$, as well as of the effective finite size parameter 
$C_{fin}$.
{From this variation, we study their influence on nuclear masses, isolate the best parameters from their ability to reproduce the experimental
nuclear masses, and determine the correlations among the parameters within the constraints of the physical masses.}

To reasonably restrict the huge space of the multi-dimensional prior parameter distribution, we need to {know their average values and 
typical uncertainties.}
We extract a part of this information from a recent work where these data were compiled from a large number of Skyrme, Relativistic Mean Field and 
Relativistic Hartree-Fock models~\cite{Casali1}. 
For each of them, the average and the standard deviation among the model predictions were estimated, and are reported in the second line of Table~\ref{tab:ETF}. 
Alternatively, an optimized value of the empirical parameters can be estimated by a least-square fit of  nuclear masses of some chosen nuclei with the full variational ETF, with resulting parameters also given in Table~\ref{tab:ETF} (first line).
From Table~\ref{tab:ETF} we can see that the different approximations and optimization techniques lead to different values for the empirical parameters, as expected. 
The differences are however consistent with the standard deviations associated with the parameters in Ref.~\cite{Casali1}, also shown in the table (last line).

\begin{table*}[tb]
\centering
\setlength{\tabcolsep}{2pt}
\renewcommand{\arraystretch}{1.2}
\begin{ruledtabular}
\caption{Empirical parameters used in the spherical ETF model, with different approximations (first and second line). %The third line gives the variation interval around the reference value used for the prior distribution.
For comparison, the average and standard variation of the different parameters recommended in 
Ref.~\cite{Casali1} is also given. 
}
\label{tab:ETF}
\begin{tabular}{ccccccccc}
%\hline
  Parameter & $ n_{sat}$ & $E_{sat}$ & $K_{sat}$ & $E_{sym}$ & $L_{sym}$ & $K_{sym}$ & $m^*_{sat}/m$ & $C_{fin}$ %& $\sigma$
\\
    {} & ($fm^{-3}$) & (MeV) & (MeV) & (MeV) & (MeV) & (MeV) & {}& (MeVfm$^5$)%& (MeV/A)
\\
 \hline
Full ETF, optimized set & 0.1589 & -15.84 & 266.73 & 32.81 & 58.37 & -37.87 & 0.7484 & 62.15 %&\cred{XXX}
\\
\hline
 Analytical ETF, reference set & 0.1540 & -16.04 & 255.91 & 33.43 & 77.92 & -2.19 & 0.7  & 59%&\cred{XXX}
\\
variation interval (absolute) & $\pm$0.0051 & $\pm$0.20 & $\pm$34.39 & $\pm$2.64 & $\pm$30.84 & $\pm$142.71 & $\pm$0.15&$\pm$13\\
variation interval (relative)  & $\pm 3.31 \%$ & $\pm 1.25\%$ & $\pm 13.44\%$ & $\pm 7.90\%$ & $\pm 39.58\%$ & $\pm 6516.44\%$ & $\pm 2.14\%$ & $\pm 22.03\%$ \\

\hline
Average and standard deviation from Ref.~\cite{Casali1} & 0.155 & -15.8 &  230 & 32 & 60 & -100 & 0.75 \\ & $\pm$0.005 & $\pm$0.3 & $\pm$20 & $\pm$2 & $\pm$15 & $\pm$100 & $\pm$0.1 \\ 
%\hline
\end{tabular}
\end{ruledtabular}
%\end{minipage}
\end{table*}

It is important to stress that, because of the possible correlations among empirical parameters, the set corresponding to the average values (second line of Table ~\ref{tab:ETF}) is not necessarily an optimized set, but only represents the central value of our prior parameter distribution. Still, we can use it as a reference set to determine a reasonable domain for the extra parameter $C_{fin}$ which is specific to our meta-modeling and cannot therefore be taken from the literature.

\begin{figure}[htbp]
    \begin{center}
\includegraphics[width=0.35\textwidth,angle=270]{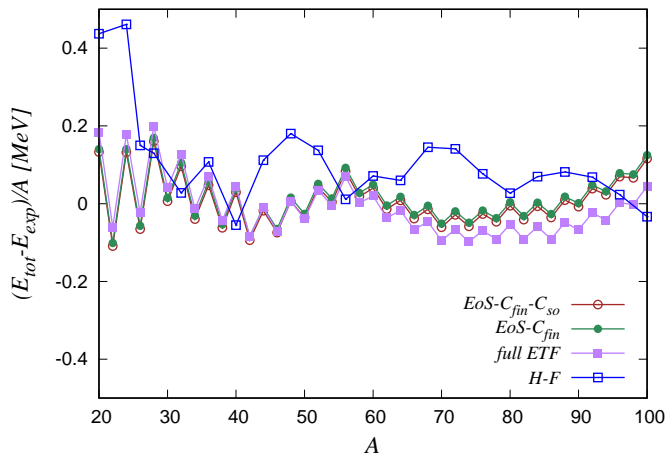}
      \caption{(Color online) Difference between theoretical and experimental  energy per particle of symmetric nuclei for the reference analytical EoS,
      in the case where a single surface parameter $C_{fin}$ is added to the empirical set (filled circles), and in the case where a spin-orbit
      coupling $C_{so}$ is also added (open circles).  
      The result for the full ETF using an optimized EoS (filled squares) and spherical Hartree-Fock results using the Sly4 \cite{Sly4}
parameter 
      set (open squares) are also given.
      }\label{fig:aabdiff_eosopt}
      \end{center}
\end{figure}

We optimize $C_{fin}$ by calculating the binding energy of symmetric nuclei with the analytical mass model, using for the other parameters the average empirical set from Table~\ref{tab:ETF}. 
We minimize with respect to $C_{fin}$ the average relative dispersion 
$\sum_{i=1}^N (E_{i,th}-E_{i,exp})^2/E_{i,exp}^2,$
for the total number $N$ of available nuclear masses, 
where the theoretical values $E_{th}$ are calculated 
from the analytical ETF model Eq.~(\ref{eq:binding}), and the experimental values $E_{exp}$ are taken from the AME2012 mass table~\cite{AME2012}.
The resulting value is $C_{fin}=59$ MeV fm$^5$. We call this parameter set as EoS-$C_{fin}$, as displayed in Fig.~\ref{fig:aabdiff_eosopt}.

In realistic Skyrme functionals, other energy terms (spin-orbit, spin-gradient, isovector gradient terms) are added in addition to the surface term.  To check whether a single effective isoscalar finite size term is sufficient to catch the information contained in nuclear masses, we introduce an additional spin-orbit term with coefficient $C_{so}$ to the ETF functional Eq.~(\ref{eq:binding}), and minimize the average relative dispersion in the two-dimensional 
$(C_{fin}, C_{so})$ space, see results EoS-$C_{fin}$-$C_{so}$ in Fig.~\ref{fig:aabdiff_eosopt}. 
Though the optimal value of $C_{fin}$ is obviously modified by the presence of the spin-orbit term (from $C_{fin}=59$ to $C_{fin}=61$ MeV fm$^5$) , the  quality of data reproduction is not modified as we can see from Fig.~\ref{fig:aabdiff_eosopt}. This shows that this extra parameter is redundant as far as binding energies are concerned.

We can observe from Fig.~\ref{fig:aabdiff_eosopt} that  the deviation of the ETF model from the experimental data is much higher than in full optimized DFT calculations, {see for instance the best DFT in Ref.~\cite{goriely} where a residual difference between the DFT and the experimental total masses are of the order to 500~keV. Notice however that this excellent comparison is possible by introducing empirical corrections to the theoretical mass model.
Without these corrections, the differences are larger and of the order of a few MeV.
For reference, we show in Fig.~\ref{fig:aabdiff_eosopt} the comparison for the Skyrme Hartree-Fock (H-F) SLy4 mean field model, as well as the result of a full variational determination of the empirical parameters using the complete ETF with the empirical EoS, plus the extra $C_{fin}$ term (see section \ref{sec:full_ETF}). We can see that both calculations show a performance comparable to the one of the analytical mass model. 
This means that the observed deviation is not due to the limitations of the analytical mass model, but rather to the fact that we are working in the mean-field approximation and supposing spherical symmetry. 

{This comparison shows that our present ETF mass model is able to reproduce satisfactory well the nuclear masses, averaging over shell corrections, and can be used further for extracting constraints and correlations
among the empirical parameters.}

The systematic error of the meta-modeling on symmetric nuclei can be very roughly 
estimated as $\Delta E \approx 100 \times A$ keV.

In our simplified functional Eq.(\ref{eq_sym_density_energy_Skyrme}) the finite size term is purely isoscalar, while standard Skyrme functionals contain also isovector couplings. 
In order to test if such a term is needed for a reasonable reproduction of nuclear masses,
we calculate the residuals for a large number of isotopic chains for Z=20, 28, 50, 82, using the $C_{fin}=59$ MeV fm$^5$ value extracted from the analysis of symmetric nuclei. 
The results are displayed
in Fig.~\ref{fig:gasymbdiff_varyz}  and show that the residuals are of the same order for all asymmetries $I$. 
Again, the reproduction of experimental data is of comparable quality if the full ETF functional is used without approximations, using  the optimized parameter set of Table \ref{tab:ETF} within the empirical model (filled squares). Spherical HF calculations for even-even nuclei within the SLy4 parametrization \cite{Sly4} (empty squares) are also included in the figure. The presence of shell effects modifies the global shape of the residuals, but again the absolute value of the deviation is comparable.
In the case of the full ETF, no surface isovector term is included, while this latter is taken into account in the SLy4 calculation.

We can conclude that the analytical ETF mass model leads to 
a systematic error on binding energies of the order of $\Delta E \approx 100 \times A$ keV, independent of the neutron richness. 
The introduction of an isovector surface parameter would not improve the predictive power of the meta-modeling, and a single isoscalar parameter is sufficient to reproduce the masses even along the full isotopic chains, within the precision allowed by the classical spherical ETF approximation,
as it was already suggested in Ref.~\cite{bulgac15}.

This somewhat surprising result might be due to the fact that within the ETF formalism the isovector surface energy depends in a non-trivial way on the EoS parameters and diffusivities, and 
is non-zero even without an isovector surface coupling  \cite{Francois2}.
\\

%% Figure in four panels

\begin{figure}[htbp]
    \begin{center}
\includegraphics[width=0.3\textwidth,angle=270]{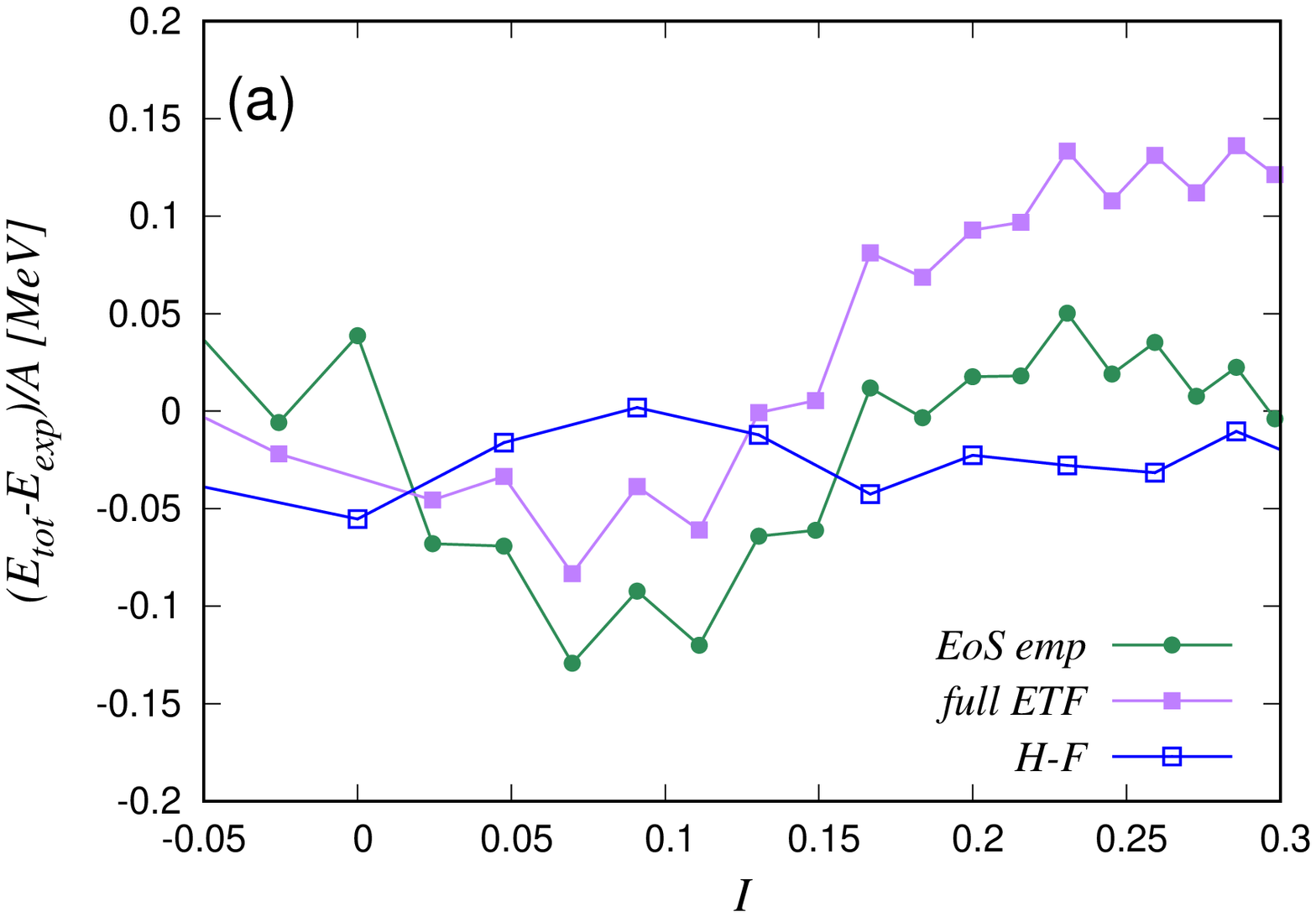}
\includegraphics[width=0.3\textwidth,angle=270]{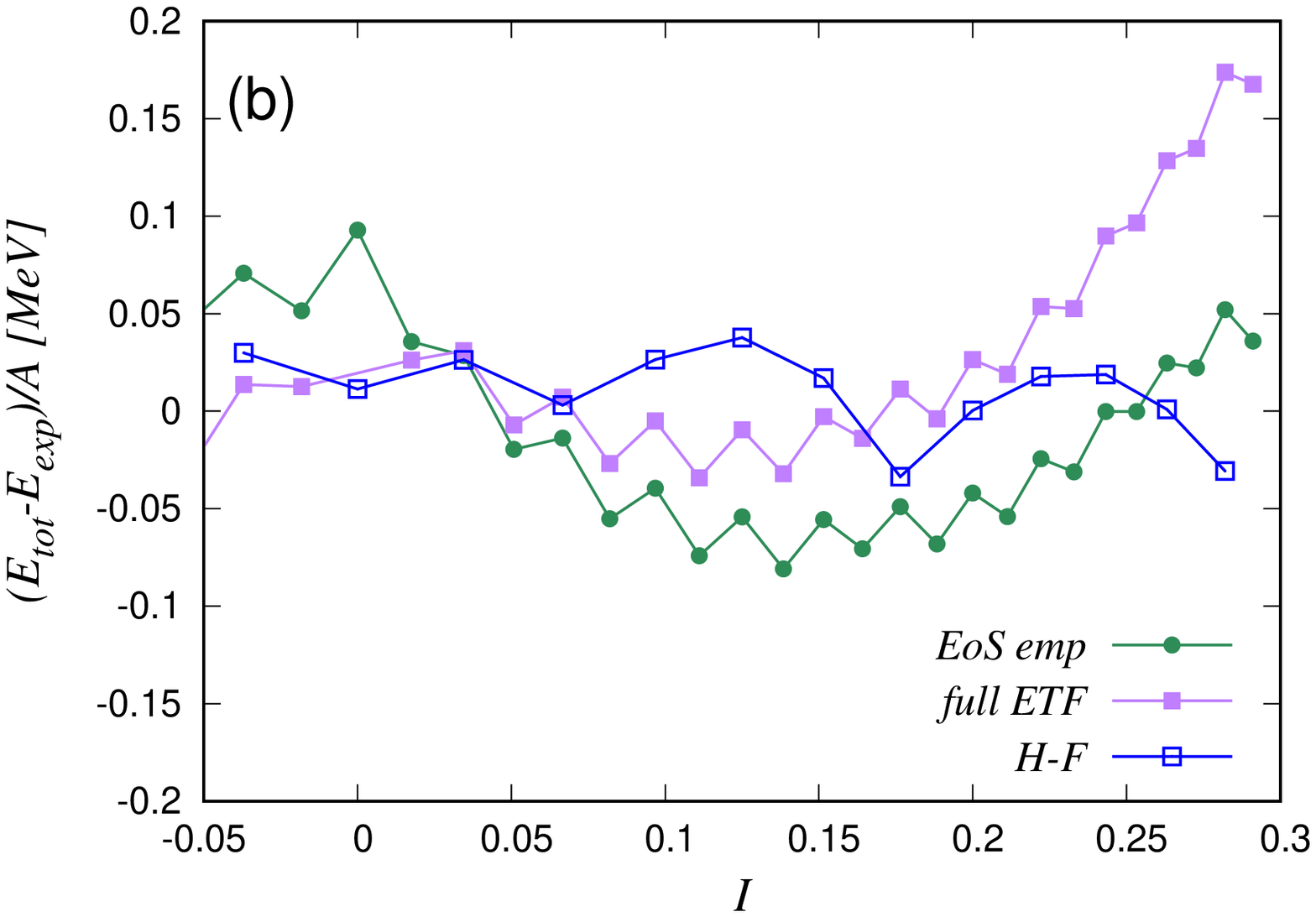}
\includegraphics[width=0.3\textwidth,angle=270]{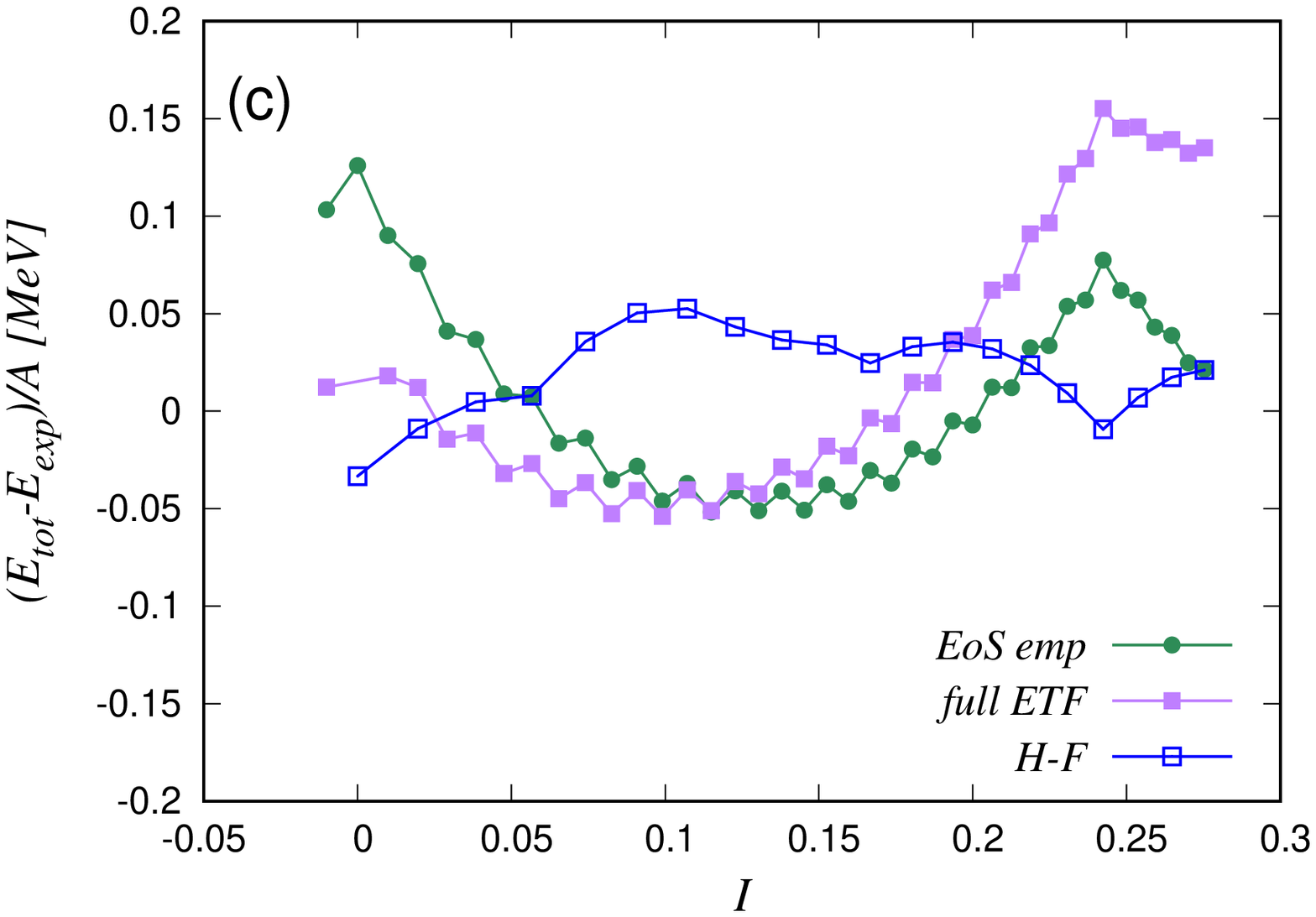}
\includegraphics[width=0.3\textwidth,angle=270]{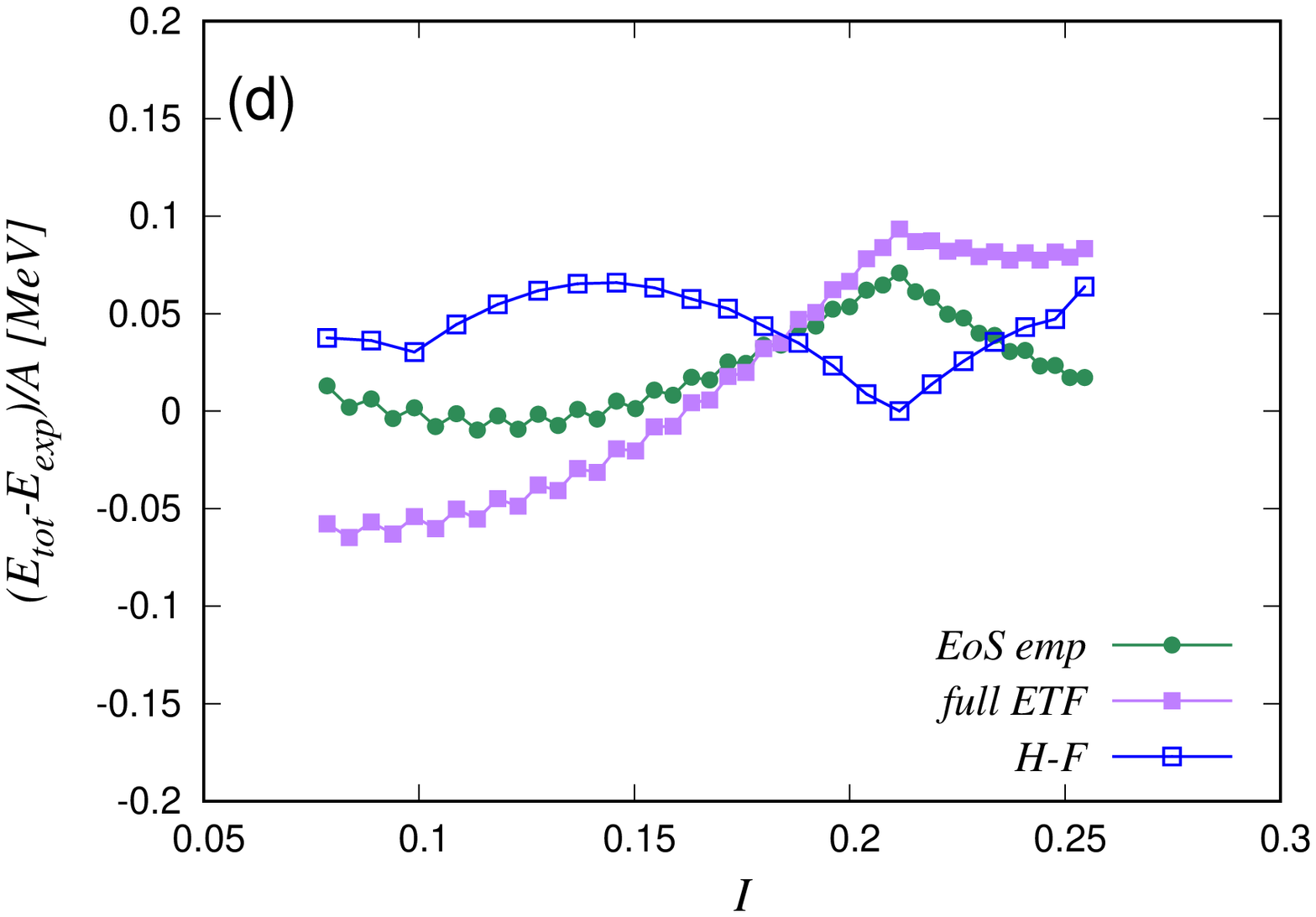}
      \caption{(Color online) Difference between theoretical and experimental energy per particle vs asymmetry
 $I=(N-Z)/A$ for different $Z$ values (a) 20, (b) 28, (c) 50 and (d) 82, for the reference empirical EoS (filled circles). The result for the full ETF 
mass model using an optimized EoS (filled squares) and spherical H-F results with the Sly4 parameter set (open squares) are also given.
      }\label{fig:gasymbdiff_varyz}
      \end{center}
\end{figure}

The width of the prior distribution for  the finite size parameter $C_{fin}$ is estimated by varying $C_{fin}$ in the analytical mass model such that the residuals
$(E_{th}-E_{exp})/A$ lie within $\pm$ 0.5 MeV, which leads to a value of $\Delta C_{fin}\approx 13$ MeV fm$^5$. %(see Fig. \ref{fig:aabdiff_z50_empopt_cfinsig}).
Concerning the set $\{P_\alpha\}$ of the EoS parameters, we vary them in the range $\langle{P}_\alpha\rangle -\sigma_\alpha \le P_\alpha \le \langle{P}_\alpha\rangle +\sigma_\alpha$, with averages and standard deviations from Table~\ref{tab:ETF}.

%\newpage  

To isolate the influence of the other empirical parameters $ \{P_\alpha\}$ on the nuclear mass, we now vary each parameter individually %within their standard deviation
and calculate the energy residuals for different symmetric nuclei and for the Sn isotopic chain. 
The summary of the effect on the energy residuals of variations of the empirical parameters within error bars,
is displayed in Fig.~\ref{fig:aabdiff_sig}.
Evidently, changing the isovector parameters $E_{sym},L_{sym},K_{sym}$ does not affect the residuals for the symmetric case (I=0). Therefore we plot only the influence of the isoscalar parameters on the energy residuals.% for symmetric nuclei.
 We can see from Fig.~\ref{fig:aabdiff_sig} that all isoscalar parameters are individually correlated to the nuclear mass, and the largest effect is from $m^*$. As expected, the higher the isospin ratio $I$, the larger is the influence of isovector parameters. 
Very similar results are obtained, if the full ETF mass model is used, with variationally determined parameters for the density profile and consistent inclusion of the Coulomb energy in the variation (not shown).

\begin{figure}[htbp]
    \begin{center}
\includegraphics[width=0.35\textwidth,angle=270]{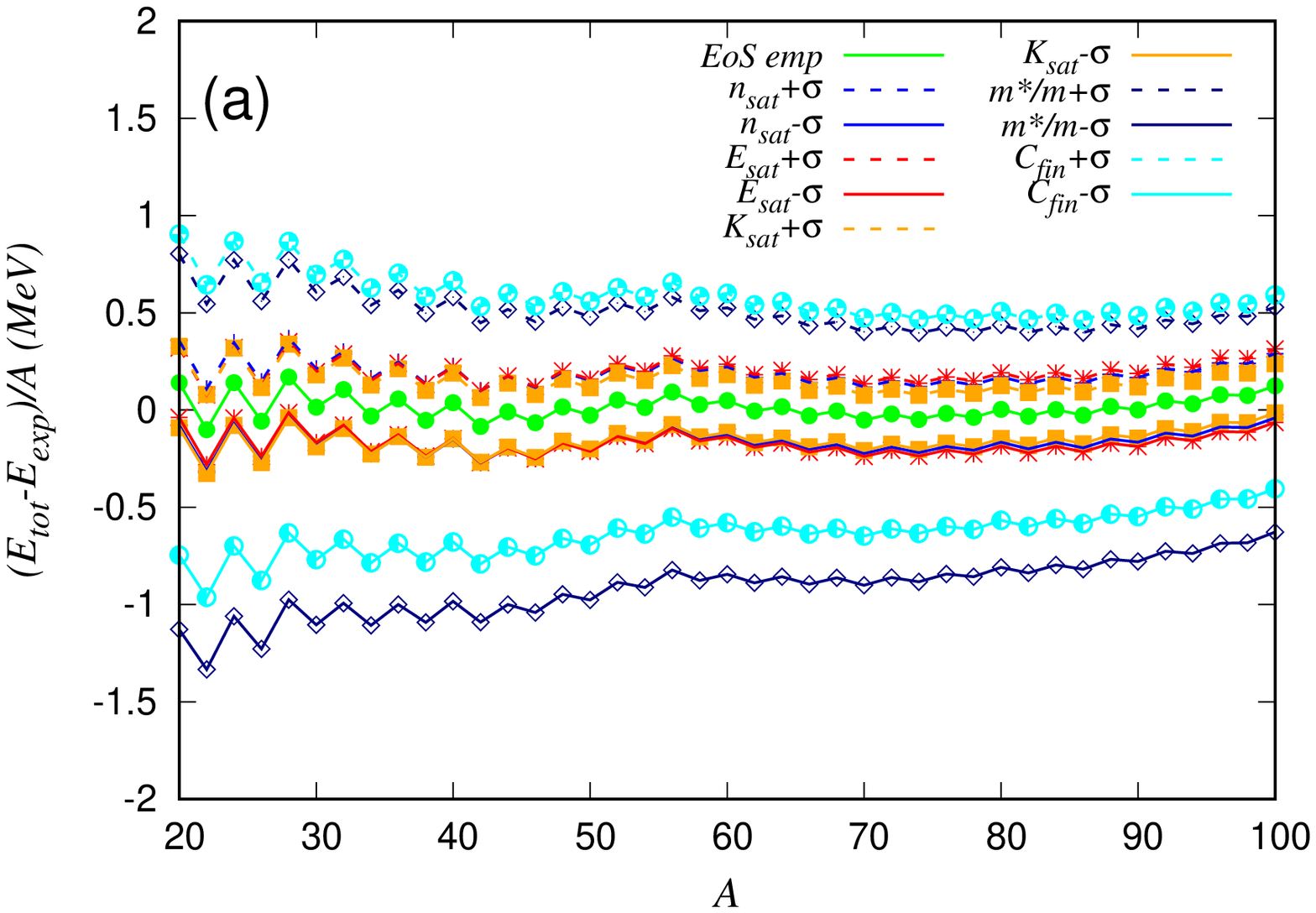}
\includegraphics[width=0.35\textwidth,angle=270]{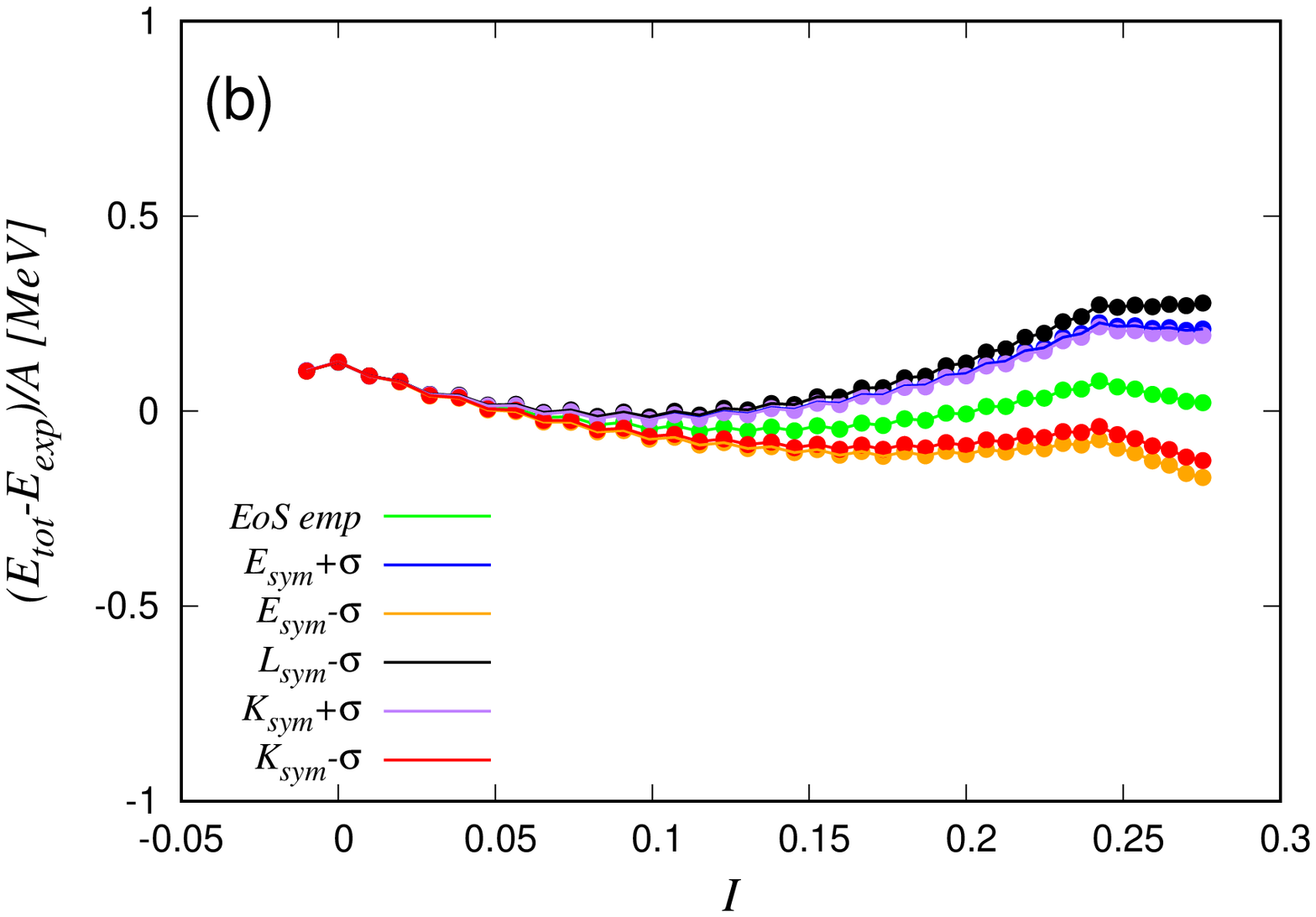}
      \caption{(Color online) Difference between theoretical and experimental binding energy per particle, for values of
      empirical parameters varied between $\pm \sigma$. 
      Panel (a): isoscalar parameters for symmetric nuclei, as a function of $A$. 
      Panel (b): isovector parameters for $Sn$ isotopic chain, as a function of $(N-Z)/A$.  
      }\label{fig:aabdiff_sig}
      \end{center}
\end{figure}

The existence of such correlations does not necessarily mean that imposing a precise reproduction of nuclear mass will allow us to restrict the uncertainty domain of the empirical parameters, because the different parameters are a-priori independent, and should therefore be independently varied. This we do in the next section.

\section{Results}

We now show the results obtained with the analytical and full ETF mass models employing the meta-functional. We recall that the parameter of the meta-functional are the set of empirical EoS parameters, complemented by the single parameter $C_{fin}$ for the finite-size effect.

\subsection{Exploring the parameter space} \label{sec:filter}

Because of the close similarity between the results obtained with the analytical mass model (see Section \ref{sec:mass_model}) and the ones using the full variational 
determination of the density profile (see Section \ref{sec:full_ETF}), in this section we only use the analytical mass model, which is computationally very fast.

We begin with the minimum bias hypothesis: we construct a prior distribution of empirical parameters in all dimensions 
by randomly picking up values within their uncertainty
as given in Table~\ref{tab:ETF}. 

Next, we ask ourselves whether the average value of the empirical parameters can be modified, and their  uncertainty domain  reduced, by the constraint of reproduction of experimental binding energies.  

The posterior distribution after application of the mass filter is given by:
\begin{equation}
p(\{P_\alpha\},C_{fin})={\cal N} w_{filter}(\{P_\alpha\},C_{fin})\prod_{\alpha=1}^7 g_\alpha(P_\alpha) g_C(C_{fin}),
\end{equation}

where ${\cal N}$ is a normalization, and the functions $g_\alpha$, $g_C$ are the priors, here taken as flat distributions  in the range $\langle{P}_\alpha\rangle -\sigma_\alpha \le P_\alpha \le \langle{P}_\alpha\rangle +\sigma_\alpha$.
The standard choice for the filter function $w_{filter}$ is given by the likelihood probability:

\begin{equation}
w_{\chi}(\{P_\alpha\},C_{fin})= \exp (-\chi^2(\{P_\alpha\}, C_{fin})/2),
\label{eq:likehood}
\end{equation}

with the $\chi^2$ function defined as:

\begin{equation}
\chi^2 = \frac{1}{N-7} \sum_{i=1}^N \left(\frac{E_{i,th}-E_{i,exp}}{100 A_i}\right)^2.
\label{eq:chi2}
\end{equation}

Here, energies are given in KeV,  and the sum extends over all the symmetric nuclei
with $10\le Z \le 50$ and all the experimentally known isotopes of the semi-magic elements with $Z=20,28,50,82$. The denominator corresponds to the systematic theoretical error, chosen such as to have 
$\chi^2_{min}\approx 1$ over the parameter space sample\cite{dobaczewski}. 
The experimental error bar on the masses is always much smaller than the systematic theoretical error, and has been neglected. 

To better visualize the effect of the filter and at the same time have a convenient representation of the average binding energy 
deviation associated with each parameter set, 
we  also introduce a dimensional quantity analogous to the $\chi^2$, which directly measures the average binding energy deviation over the symmetric nuclei as well as  semi-magic isotopic chains:
 
\begin{equation}
\Delta = \frac{1}{N}  \sum_{i=1}^{N}   \frac{\left|E_{i,th}-E_{i,exp} \right|}{A} ~,
\label{eq:delta}
\end{equation}
 
where the sum extends over the same nuclei as for Eq.(\ref{eq:chi2}).
 
We then define a filter function selecting  those parameter sets for which $\Delta < \Delta_\mathrm{cutoff}$, where $\Delta_\mathrm{cutoff}$ is varied  in order to
observe the influence on the different observables as well as model parameters:
\begin{equation}
w_{\Delta}(\{P_\alpha\},C_{fin})=\Theta ( \Delta_\mathrm{cutoff}-\Delta)
\label{eq:cutoff}
\end{equation}
The corresponding posterior distribution $p_\Delta$ thus explicitly depends on the chosen 
value of $\Delta_\mathrm{cutoff}$, and so on the goodness requested from the models to reproduce the data.

Once the sample of models is filtered according to the chosen cut-off,  
centroids and standard deviations for any observable $O$ can be calculated by integrating the posterior  distribution $p$  over the   model parameters as:
\begin{eqnarray}
\langle O \rangle &=& \int dP_1\dots dP_7 dC_{fin} O 
 p(\{P_\alpha\},C_{fin}), \\
 \sigma^2_O &=& \int dP_1\dots dP_7 dC_{fin} \left [ O 
-\langle O \rangle \right ]
 p(\{P_\alpha\},C_{fin}).
\end{eqnarray}
In these expressions, the value of the observable $O$ in the integral is evaluated within the considered parameter set, $O\equiv O(\{P_\alpha\},C_{fin})$, and $p$ can be $p_\chi$ (using Eq.~(\ref{eq:chi2})) or $p_\Delta$ (if Eq.~(\ref{eq:cutoff}) is used). We remark that $O$
may or may not coincide with one of the model parameters $\{P_\alpha\},C_{fin}$.   

 Some chosen examples using the cutoff filter $w_\Delta$, namely the binding energy per nucleon, rms neutron radius of $^{208}Pb$, and $ n_{sat}$ empirical parameter, 
 are shown in Fig.~\ref{fig:cutoff}. 
For each panel, the centroid and standard deviation of the likelihood posterior distribution are additionally given in Table~\ref{tab:centroid}.
\\

\begin{figure*}[htbp]
%\begin{center}
\vskip -2cm  
\hspace*{-2 cm}
\subfigure{ 
\includegraphics[width=.42\textwidth]{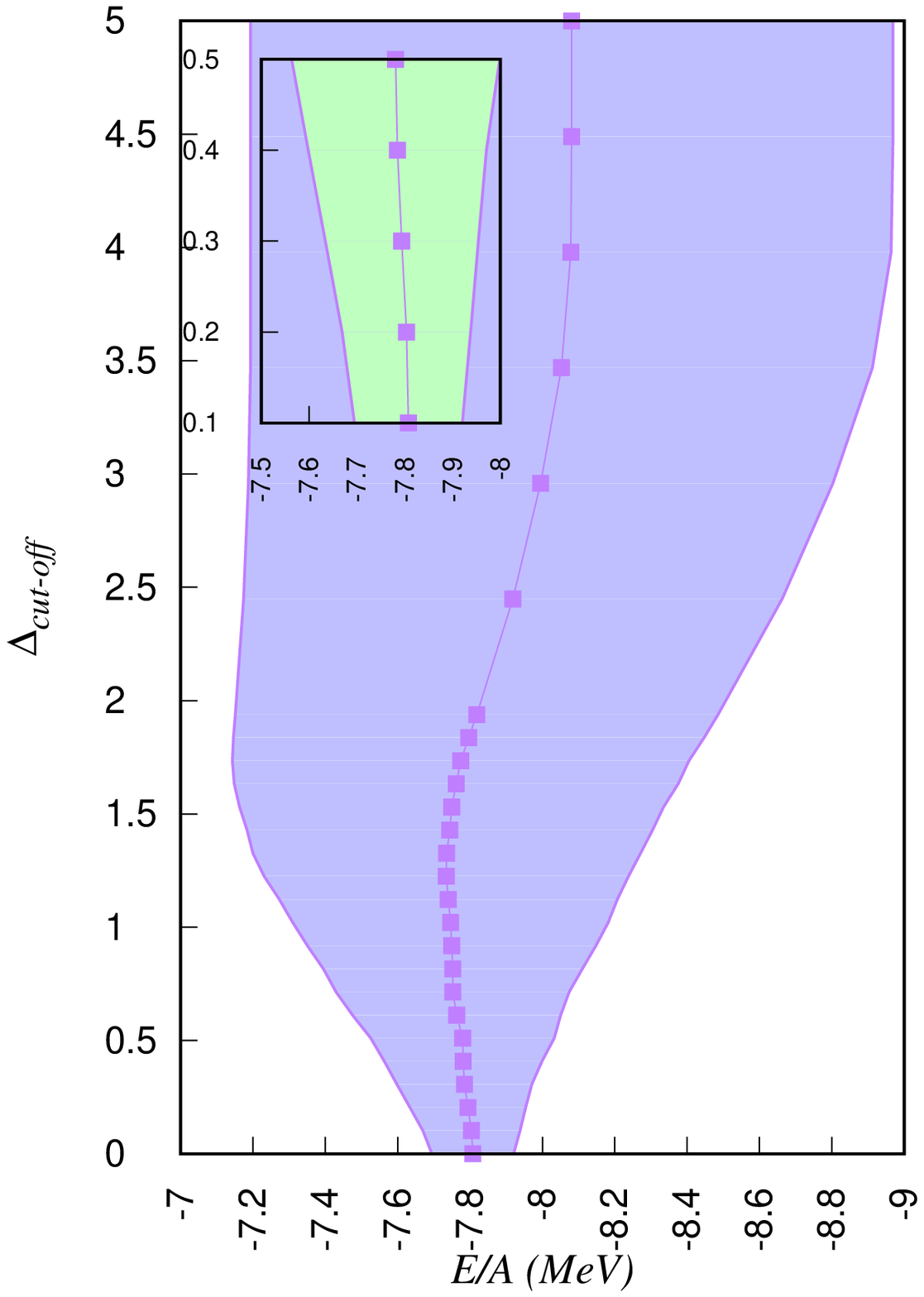}
%\vskip -2cm        
\hspace*{-1.8 cm}
      \includegraphics[width=.42\textwidth]{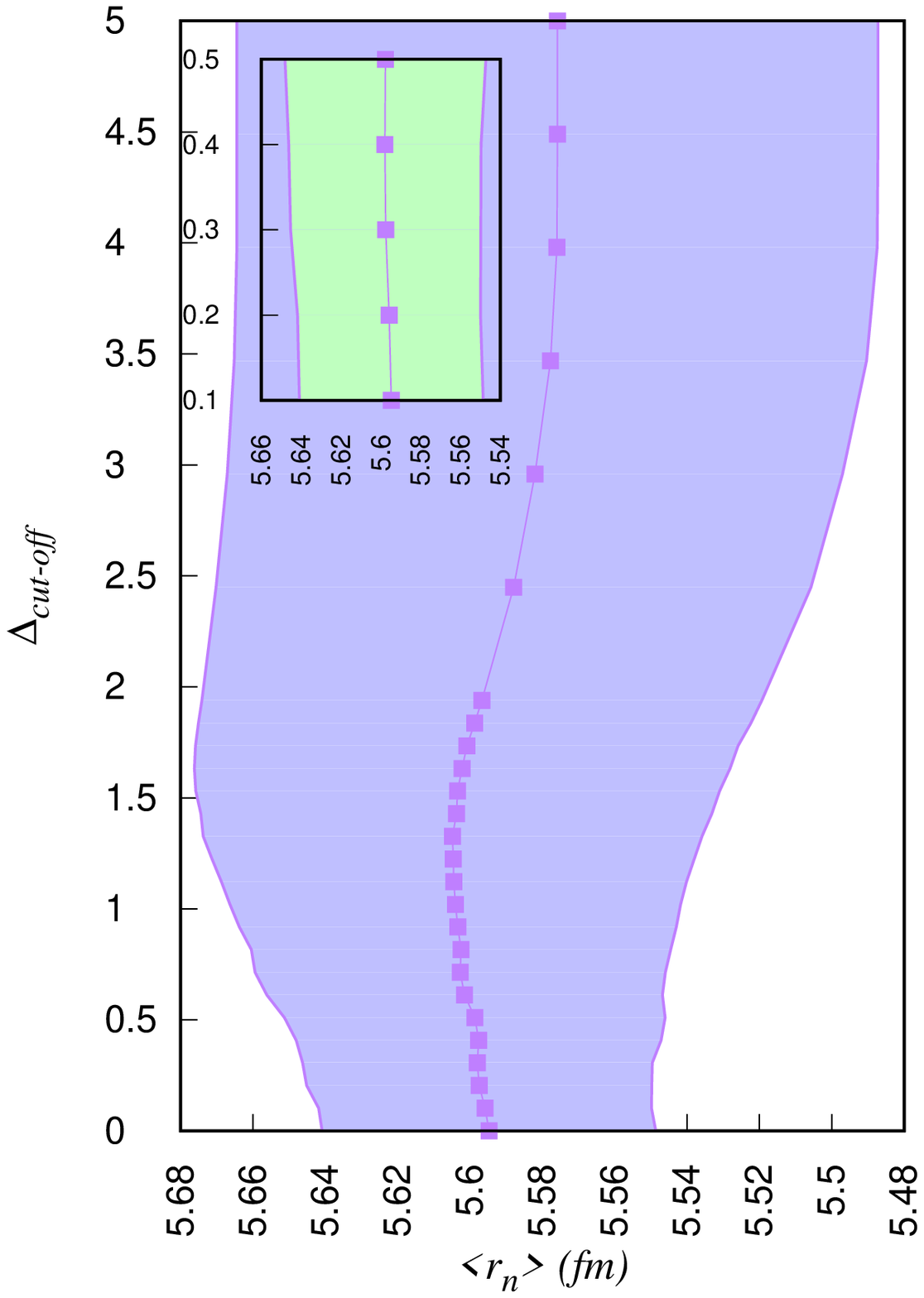}
      }
%\vskip -2.3cm  
\hspace*{-1.6 cm}
      \includegraphics[width=.42\textwidth]{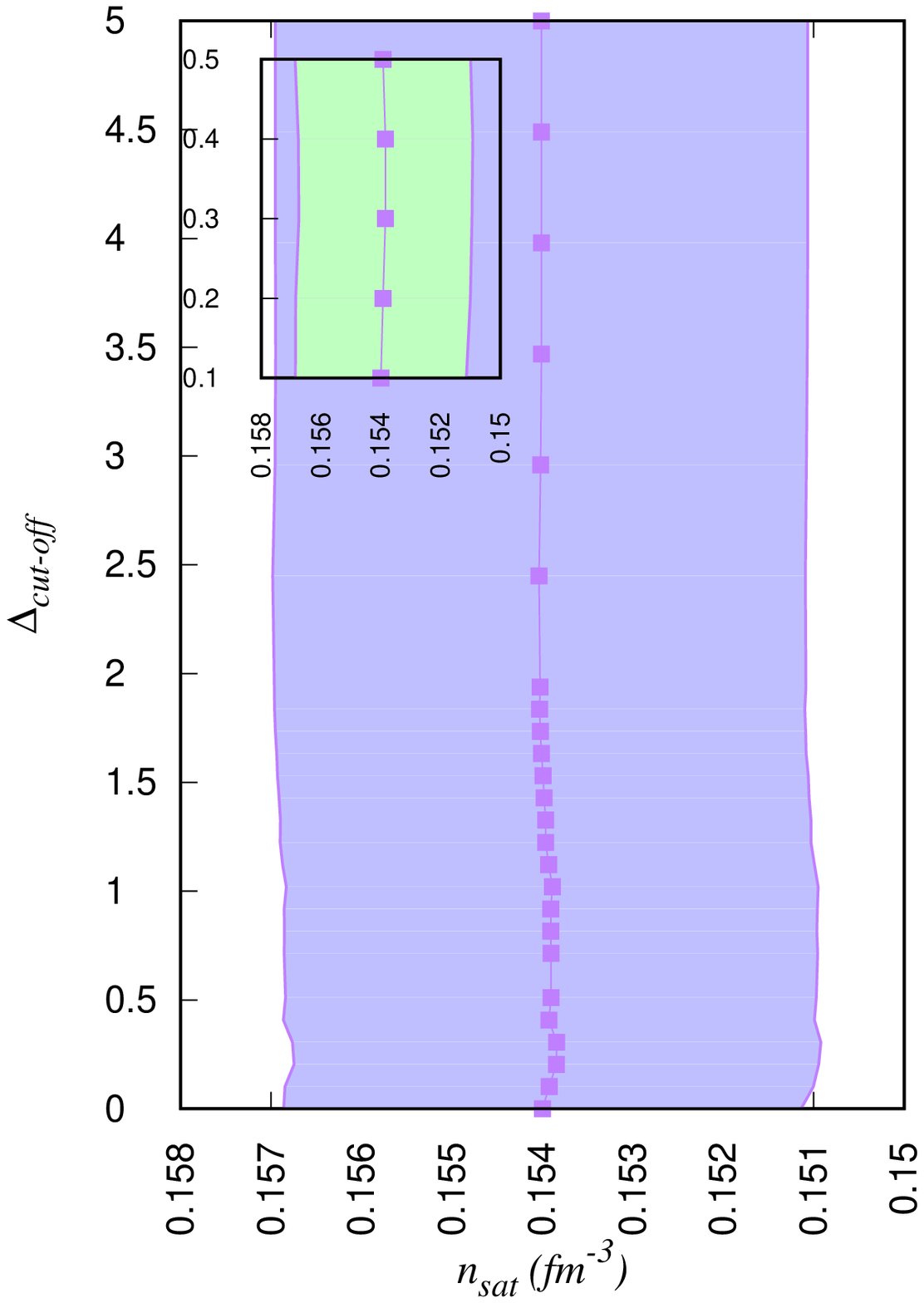} 
      
\caption{(Color online) Effect of cut-off in $\Delta$ on the average (central purple line) and standard deviation (band width) for two different observables: the $^{208}$Pb binding energy per nucleon (left) and its neutron radius (center); and the $n_{sat}$ empirical parameter (right).  In each panel, the inset shows the same, for $\Delta_\mathrm{cutoff} < 0.5$.}
      \label{fig:cutoff}
%    \end{center}
 \end{figure*}

 In each of the panels, the lines indicate the mean value $\langle O \rangle$ and the values within a standard deviation of $\sigma$.
The shown behaviors are representative of the general evolution with $\Delta_\mathrm{cutoff}$ of all the quantities we have examined. Specifically, all binding energies behave very similarly to the left panel of Fig.~\ref{fig:cutoff}; all neutron and charge radii behave as the central panel of Fig.~\ref{fig:cutoff}; and all empirical parameters show a very constant behavior as in the right panel of Fig.~\ref{fig:cutoff}.
For all observables and model parameters, the average values only slightly evolve with $\Delta_\mathrm{cutoff}$, showing that our reference set is not
far away from an optimized set.
The variance of the energy per particle monotonically decreases, showing that $\Delta_\mathrm{cutoff}$ is indeed a measure of the quality of reproduction of individual binding energies. The variance of the radii also globally decreases. This is also expected because a smaller value of $\Delta_\mathrm{cutoff}$ corresponds to a reduction of the parameter space, and therefore of the possible variation among the different models. However we can see that the cut-off is ineffective starting at around $\Delta_\mathrm{cutoff}\approx 0.5$ MeV: a reproduction of the experimental binding energy better than 500 KeV per nucleon does not improve our uncertainty on the nuclear radius.
Finally,  the constant behavior of all the empirical parameters $\{ P_\alpha\}$ is less expected. None of them (in average as well as in standard deviation)  
depends of the goodness requested in the reproduction of binding energies. 
 
This observation suggests that, imposing a reproduction of the experimental binding energies, one does not expect to be able to greatly reduce the uncertainty of the empirical parameters. 
This is due to the fact that the $\chi^2$ hypersurface is relatively flat: all EoS parameters affect more or less
in a similar way the nuclear binding energy (see Fig. \ref{fig:gasymbdiff_varyz}). 
 Since we are not supposing any a-priori correlation among the parameters, compensations can freely occur. 
This is similar to the study of neutron stars, where compensation between different empirical parameters were observed to greatly reduce the effectiveness of the filters~\cite{Casali2}.

The only model parameter that can be better constrained by a better reproduction of nuclear mass is the finite size parameter $C_{fin}$. 
This is shown in Fig.~\ref{fig:cfin}, which displays the evolution of the variance of $C_{fin}$ with the cut-off (the average value is not affected).

\begin{figure}[htbp]
      \includegraphics[width=.35\textwidth,angle=270]{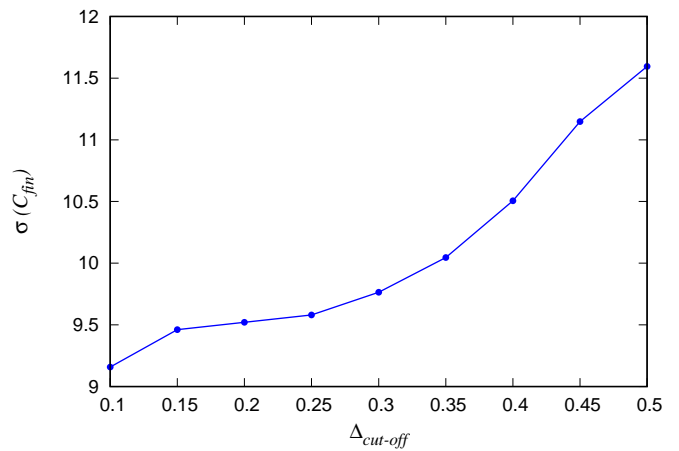}
\caption{(Color online) Variance of the posterior distribution $p_\Delta$ of the finite size $C_{fin}$ parameter as a function of the cut-off (see text). }
      \label{fig:cfin}
 \end{figure}

However, the observed reduction of the standard deviation is essentially due to the fact that we have assumed a widely spread prior for this parameter.  At a still relatively large value of the cut-off $\Delta_\mathrm{cutoff}\approx 0.25$ MeV, a convergence is observed for a non-negligible width, meaning that extreme values of $C_{fin}$ are not excluded by the binding energy reproduction, because they can be compensated by the smaller but combined opposite effect of the EoS parameters. 
 
The centroids and standard deviations obtained using the likelihood filter Eq.(\ref{eq:chi2}) instead of the cut-off dependent one Eq.~(\ref{eq:cutoff}) are given in Table~\ref{tab:centroid}. These values are almost identical to the ones obtained with the $\Delta_\mathrm{cutoff}$ filter 
for $\Delta_\mathrm{cutoff}\le 0.5$ MeV.

\begin{table}[tb]
\centering
\setlength{\tabcolsep}{2pt}
\renewcommand{\arraystretch}{1.2}
\begin{ruledtabular}
\begin{tabular}{cccc}
  Parameter & $E/A(^{208}Pb)$ & $\langle r_n \rangle(^{208}Pb)$ & $n_{sat}$  \\
  {} &  {(MeV)} & (fm) & (fm$^{-3}$) \\
 \hline
 Average & {-7.806} & {5.596} & {0.154}  \\
 \hline
Standard deviation  & {0.124} & {0.046} & {0.003} \\
\end{tabular}
\end{ruledtabular}
\caption{(Color online) Centroid and standard deviation of the likelihood posterior distribution for the observables in Fig. (\ref{fig:cutoff}).}
\label{tab:centroid}
\end{table}

To conclude, it appears from this study that an improvement in the predictive power of the mass model would not lead to any further constraint on the EoS empirical parameters beside the state-of-the-art values represented by Table~\ref{tab:ETF}.

\subsection{Correlations}

We now come to the main result of this study, namely the search for physical correlations of the different empirical parameters among themselves, as well as with the radii and skins.

\begin{figure*}[htbp]
    \begin{center}
   % \hspace*{-0.8 cm}
   % \vskip -2cm
   \hspace{-2 cm}
\includegraphics[width=0.75\textwidth,angle=270,clip]{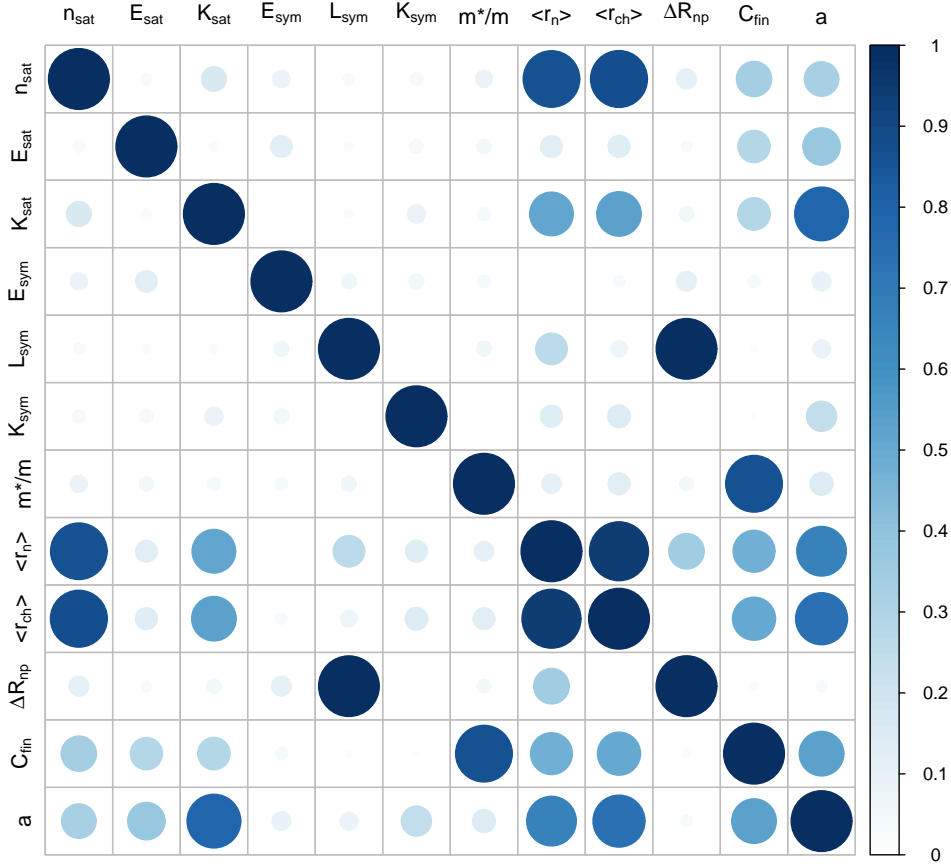}
\vskip -1cm
      \caption{(Color online) Posterior correlation matrix for empirical parameters and nuclear observables, after application of the mass filter.  
      }\label{fig:corrmatrix}
      \end{center}
\end{figure*}

The covariance study is done on the posterior distribution of empirical parameters with the binding energy filter given by Eqs.~(\ref{eq:likehood})-(\ref{eq:chi2}).
We have checked that the use of the cut-off filter produces the same correlation matrix provided   $\Delta_\mathrm{cutoff}\le0.5$ MeV.

To summarize the different correlations, we display in Fig.~\ref{fig:corrmatrix} the correlation matrix of the following quantities:
empirical parameters ($ n_{sat}$, $E_{sat}$, $K_{sat}$,$E_{sym}$, $L_{sym}$, $K_{sym}$), the effective mass $m^*/m$, the finite size parameter $C_{fin}$, and some observables in $^{208}$Pb, namely the rms charge radius $\sqrt{ \langle r^2_{ch} \rangle}$,
neutron rms radius $\sqrt {\langle r^2_n \rangle}$,  neutron skin $\Delta R_{np}$  and diffuseness $a$.  
The picture is qualitatively the same if other stable nuclei are chosen.

The first observation in this plot is that none of the empirical parameter (including the effective mass) is correlated to each other. In particular, the correlation coefficient between $L_{sym}$ and $E_{sym}$ is close to zero, at variance with numerous studies in the literature\cite{tsang_rep,KhanMargueron,Ducoin2011,Kortelainen12,Danielewicz14,McDonnel15}.
That correlation is generally explained by the fact\cite{Colo04} that nuclear structure probes the symmetry energy at densities below saturation: to have a given value for $e_{IV}(n_0<n_{sat})$, a higher (lower) value of $E_{sym}=e_{IV}(n_{sat})$ must be associated to a higher (lower) value of its derivative at $n_{sat}$, namely $L_{sym}$. However this argument neglects the effect of the second derivative $K_{sym}$. In Skyrme forces, $K_{sym}$ is typically negative and strongly a-priori correlated with $L_{sym}$, therefore the argument holds. But if we allow a larger exploration of the $K_{sym}$ parameter space, we can see that the new behaviours explored for the symmetry energy are still compatible with the binding energy constraint, and break the simple correlation between $E_{sym}$ and $L_{sym}$. 
This suggests that this commonly observed correlation is not a direct consequence of the constraint of mass reproduction, but is rather due to the lack of flexibility of the Skyrme functional.

If we now turn to the correlations between observables and empirical parameters,
apart from the trivial correlation between the neutron and charge rms radii, the strongest correlations visible in the plot are the correlations of the
radii with the saturation density $n_{sat}$, and that between the neutron skin $\Delta R_{np}$ and the slope of symmetry energy $L_{sym}$. 

The correlation of the radii with the saturation density $ n_{sat}$ is relatively trivial and expected, since the value of $n_{sat}$ determines the average density of nucleons per volume.
The constant $n_{bulk}$ is indeed a function of the saturation density in our model and it explicitly 
enters in the definition of the radii. At low $\delta$ values, the second important quantity entering in the determination of the bulk density is $K_{sat}$, see Eq.(\ref{rho0}), which explains the weaker correlation between the radii and $K_{sat}$.

The excellent correlation of $\Delta R_{np}$ with $L_{sym}$ is in agreement with previous studies in the literature~\cite{Mondal,warda09,centelles10,reinhard16,centelles13}, 
which used specific Skyrme or RMF energy functionals. 
However, other correlations shown in the literature with $E_{sym}$ or $K_{sym}$~\cite{Nazarewicz14,papazoglou} are not observed here, which might again indicate the model dependence of these correlations.

The $\Delta R_{np}$-$L_{sym}$ correlation can be understood from the fact that the skin is proportional to the average displacement between neutrons and protons $\Delta R_{HS}$ \cite{meyers}, as it can be seen from Eq.~(\ref{eq:radius}). 
Now, $\Delta R_{HS}$ is directly linked to the isospin dependence of the saturation density, see Eq.~(\ref{eq:rhs}). 
This latter is determined by the ratio $L_{sym}/(K_{sat} + K_{sym}\delta^2)$, see Eq.~(\ref{rho0}).
The value of the bulk asymmetry $\delta$ is small in stable nuclei $\delta\approx 0.1$ and the parameter $K_{sat}$ is relatively well constrained by the iso-scalar giant resonance mode. 
As a consequence, $L_{sym}$ is the key parameter to determine the neutron skin. 

A word of caution has to be given here. In our analytical mass model we have employed the approximation $a_n=a_p$. Since the diffusivity depends in a complicated way both on the empirical parameters and on the finite size ones, this simplification might lead to an overestimation of the quality of the correlation. 
%is in progress. 
Still, it should be stressed that this same excellent correlation was observed also in theoretical analysis where this approximation was not done \cite{reinhard16}.  Also, an analysis with the full ETF mass model, where $a_n(N,Z)$ and $a_p(N,Z)$ are independent parameters variationally determined for each nucleus, confirms the correlation between  
$\Delta R_{np}$ and $L_{sym}$.

Finally,  the parameters related to the nuclear surface exhibit different, though weak, interesting correlations.  
The finite size parameter $C_{fin}$ is correlated to the effective mass and, in a weaker way, to the radii.
This is consistent with our observation that $C_{fin}$ and $m^*$ are the two most influential parameters in the determination of the binding energy.
A clear correlation is observed between $K_{sat}$ and $a$. 
This is an interesting feature since these two quantities contribute to the surface energy: $K_{sat}$ represent the bulk contribution while $a$ is more complex.
The parameter $a$ is also found to be correlated with the radii $r_{ch}$ and $r_n$, as is expected.
Finally, a weak correlation is observed between $a$ and $C_{fin}$.
This reveals the complex structure of the non-linear terms in the ETF which depend on these parameters in a non trivial way.
In summary, surface terms induce some interesting correlations, but these correlations are weaker, and might be less robust, than the dominant correlation $\Delta R_{np}$-$L_{sym}$. Indeed the introduction of more gradient couplings and/or the introduction of an isovector diffuseness might change the properties of the surface parameters $a, C_{fin}$ and their mutual correlations.

\subsection{Radii and skins} \label{sec:radii}

\begin{figure}[htbp]
    \begin{center}
\includegraphics[width=0.3\textwidth,angle=270]{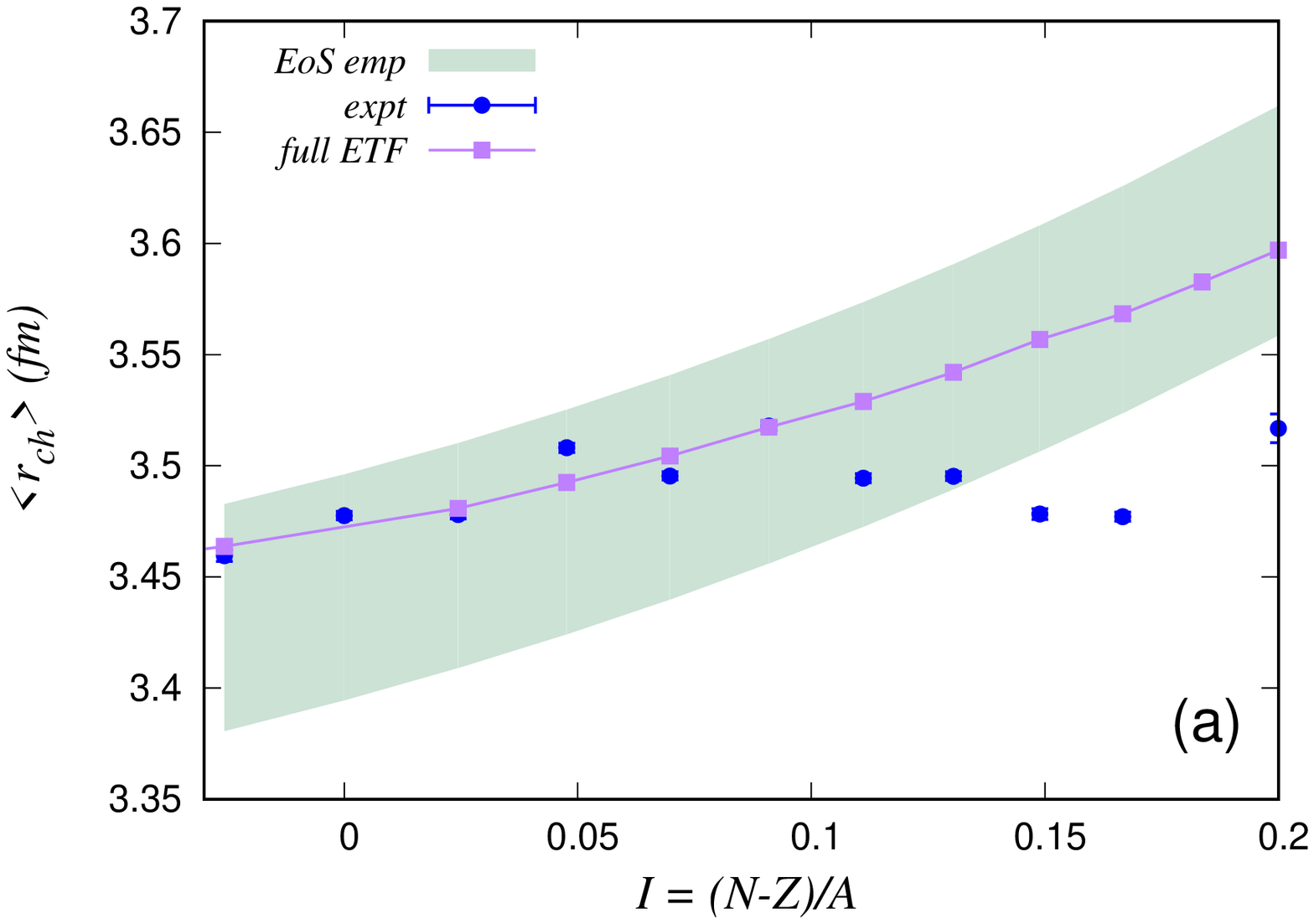}
\includegraphics[width=0.3\textwidth,angle=270]{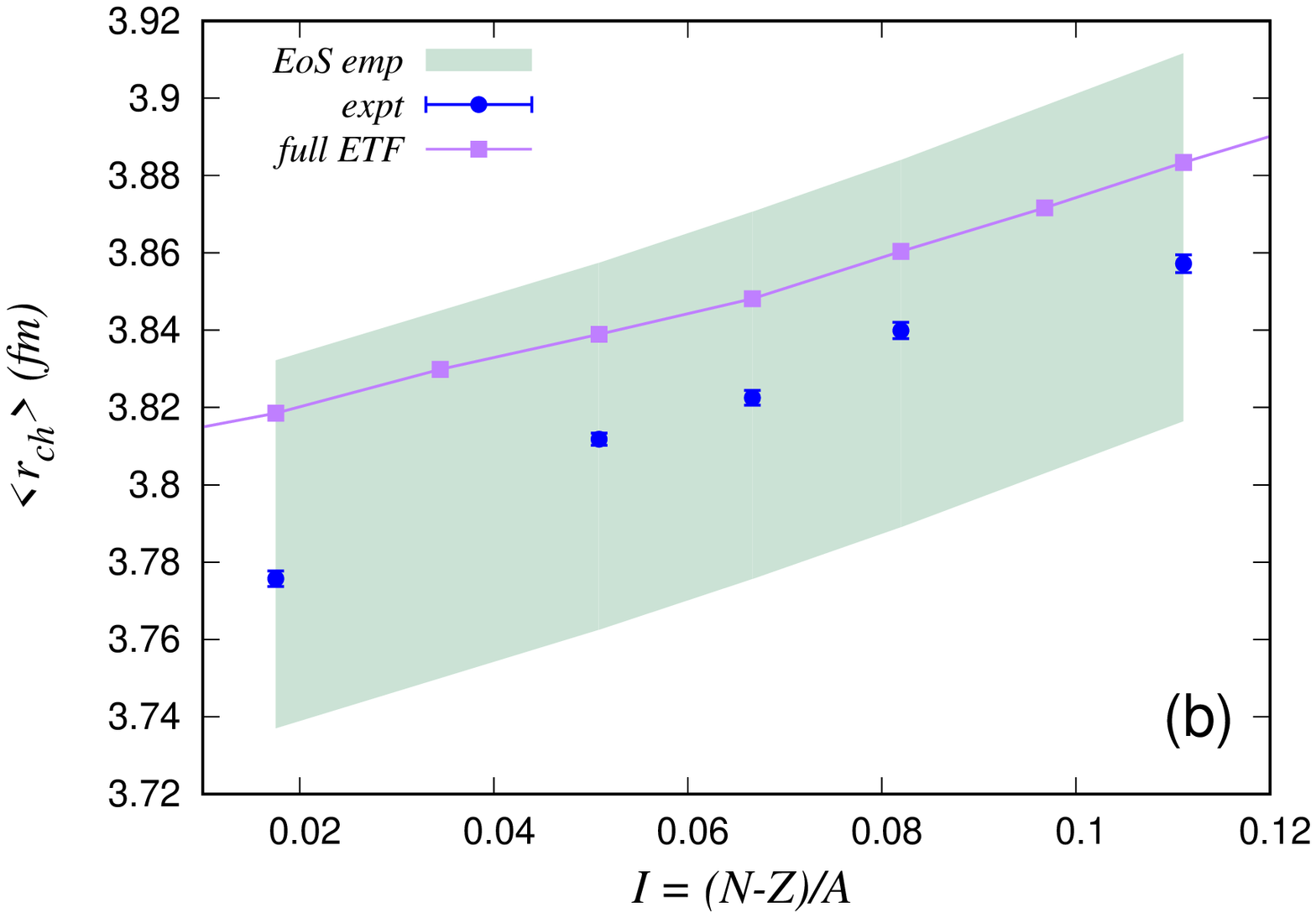}
\includegraphics[width=0.3\textwidth,angle=270]{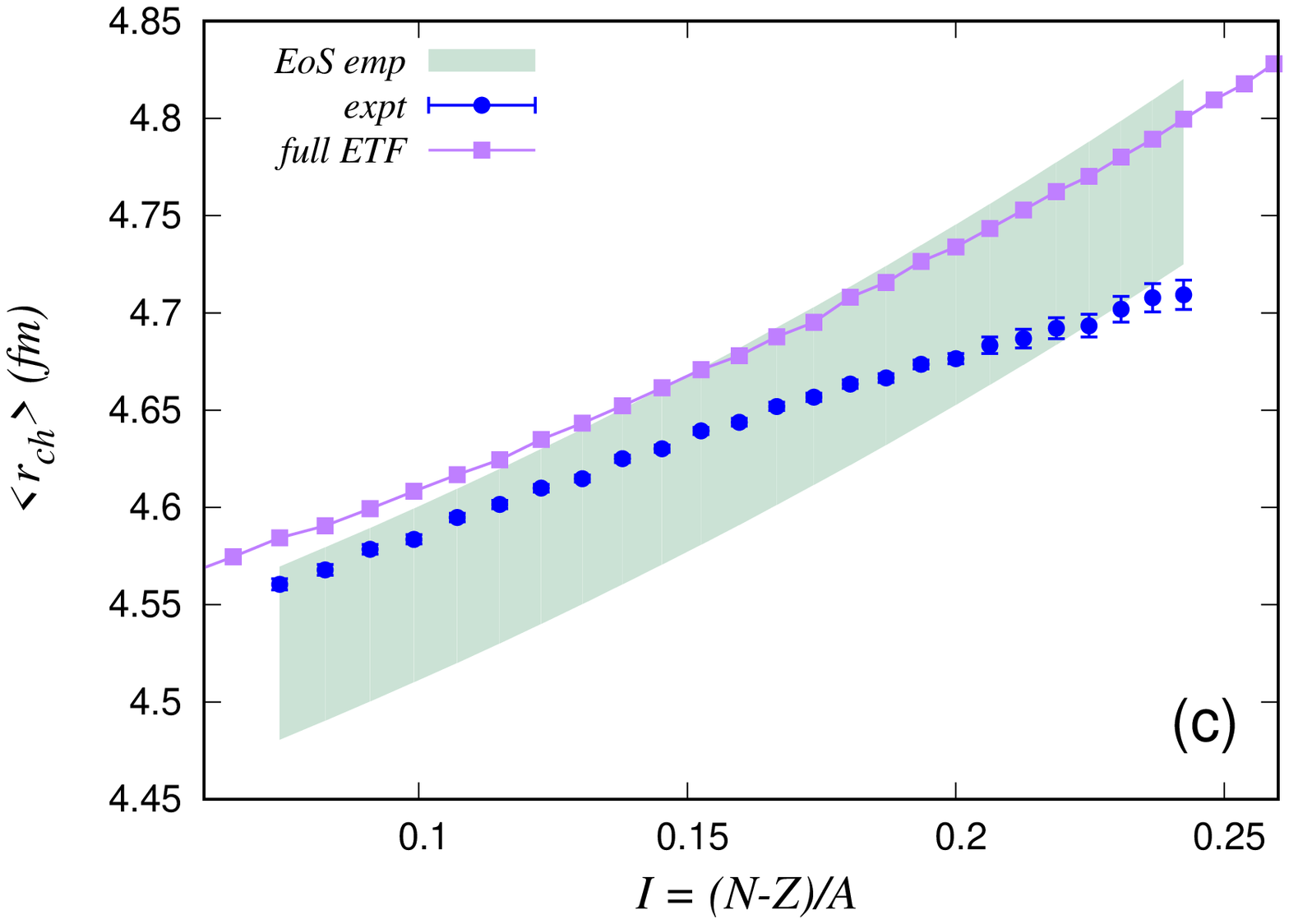}
\includegraphics[width=0.3\textwidth,angle=270]{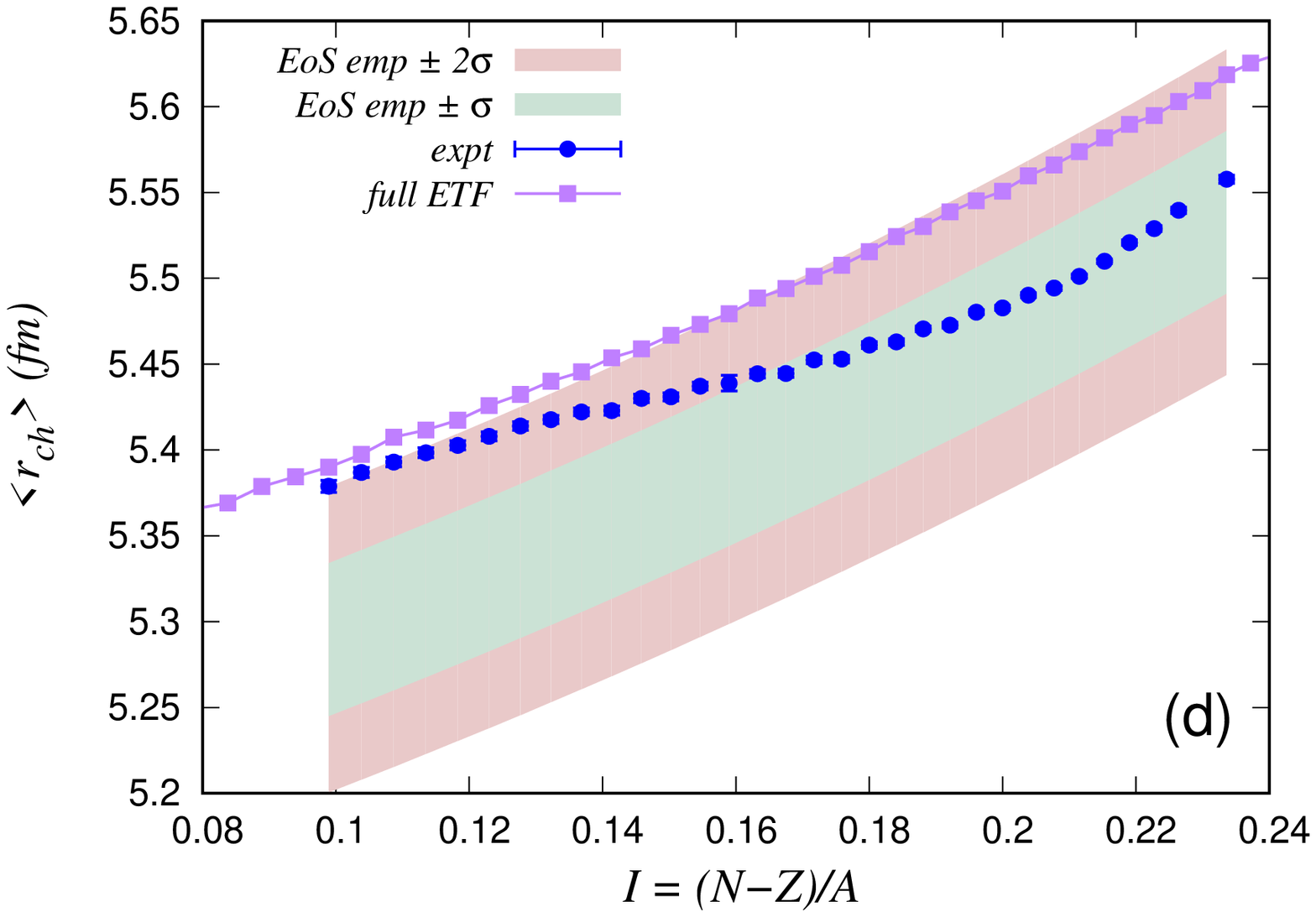}
      \caption{(Color online) Rms charge radii vs $I$ for semi-magic isotopic chains for different $Z$ values (a) 20, (b) 28, (c) 50 and (d) 82. Symbols with error bars: experimental data. 
      Bands: predictions at $1\sigma$ (and $2\sigma$) for the empirical EoS filtered through the binding energy constraint. 
      Lines with filled squares: prediction from the full variational ETF with the optimized empirical parameter set.   
      }\label{fig:aarmsch_z50_empopt_cfinsig}
      \end{center}
\end{figure}

% \newpage

We have seen in the previous section that constraints on $K_{sat}$ and $n_{sat}$ might come from the reproduction of nuclear radii, and constraints on $L_{sym}$ could be obtained from the measurement of neutron skin. 
We therefore turn to examine the predictive power of our model on these observables in the present section. 

The prediction of charge radii along the different semi-magic isotopic chains, obtained from the  models filtered with the constraint of   binding energy reproduction according 
to Eq.~(\ref{eq:cutoff}) (see Section~\ref{sec:filter}) is compared to experimental data from Ref.~\cite{Marinova} in Fig.~\ref{fig:aarmsch_z50_empopt_cfinsig}. 
 In this figure, the experimental error bars are smaller than the size of the points.
The predictions for the full ETF mass model with  empirical parameters optimized on the binding energies (Table \ref{tab:ETF}) is also given, and seen to be compatible 
at the $2\sigma$ level with the analytical mass model.
 
We can see that the reproduction is not optimal, but the performance of the model is comparable to the one of complete   ETF or DFT calculations 
in the absence of deformation~\cite{Buchinger,Patyk}.

\begin{figure}[htbp]
    \begin{center}
\includegraphics[width=0.35\textwidth,angle=270]{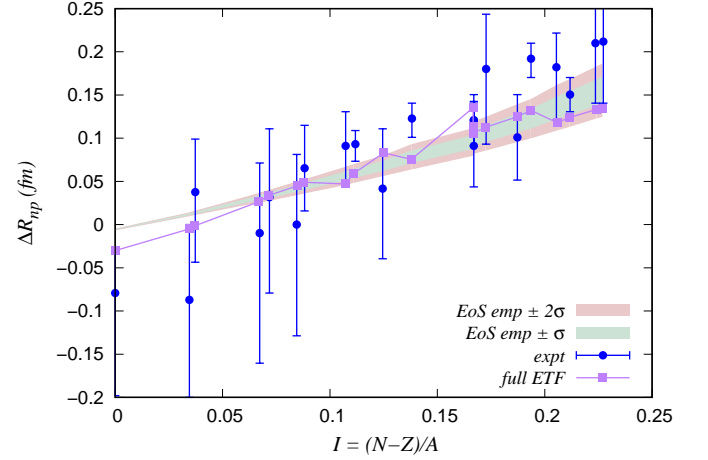}
      \caption{(Color online) Neutron skin as a function of global asymmetry for various nuclei. 
      Symbols with error bars: experimental data. 
      Bands: predictions at $1\sigma$ (and $2\sigma$) for the empirical EoS filtered through binding the energy constraint.
      Lines with filled squares: prediction from the full 
      variational ETF with the optimized empirical parameter set.     
      }\label{fig:gasymskin_fit}
      \end{center}
\end{figure}

Since the experimental error bars are much smaller than the theoretical ones, the charge radii constitute a very promising observable to further constrain the EoS. 
This is especially interesting for the $K_{sat}$ parameter, for which some tension still exists\cite{Casali1} between constraints extracted from relativistic  and non-relativistic models
\footnote{For the reproduction of the same GMR observables, the non-relativistic models prefer $K_{sat}\approx 210-240$ MeV, while higher values $K_{sat}\approx 250-270$ MeV are extracted from RMF calculations.}.
It is highly probable that this tension might be induced by the different correlations between $K_{sat}$ and $K_{sym}$ in the two families of models~\cite{Colo04,Colo08}, and could therefore be solved using the empirical EoS.

We have not attempted to put a further filter on the radii in the present study, because we consider that the present meta-modeling is not 
sufficiently sophisticated for this purpose. Indeed, we can see from Fig.\ref{fig:corrmatrix} that the radii also crucially depend, 
beside the EoS parameters, on the surface properties of the model (here: $a$, $C_{fin}$), which are treated in a simplified way in the present work. Moreover, the spherical approximation employed in this work is inadequate for many of the isotopes shown in Fig.\ref{fig:aarmsch_z50_empopt_cfinsig}. Finally, the small but systematic difference between  the results of the analytical mass model and the full ETF with increasing nuclear charge, suggests that it might be important to consistently include the Coulomb field in the variational theory, for a correct description of charge radii. For these reasons, we leave the quantitative study of extracting EoS parameters from charge radii to a future work.

Next, we study the dependence of neutron skin $\Delta R_{np}$ on the global asymmetry parameter 
$I = (N-Z)/A$, for nuclei for which experimental measurements of the neutron skin exists \cite{Trzcinska}. 
We display the results in Fig.~\ref{fig:gasymskin_fit}, along with the results from the full ETF calculation
with the optimized empirical data set of Table \ref{tab:ETF}.  
It is clear from the figure that within the uncertainties of the empirical parameters, our model predicts
neutron skins compatible with experimental results, with a comparable level of precision as complete ETF calculations~\cite{papazoglou}.
 
This shows that a different diffuseness for the proton and neutron distribution is not necessary to reproduce the correct  magnitude of the neutron skin, in contradiction with the results of 
Refs.~\cite{warda09,centelles10,centelles13}. 
We might understand this contradiction from the fact that our prescription for the diffuseness Eq.~(\ref{eq:diffuseness}) effectively contains local and non-local isovector terms in a complex way,
even assuming $a_n=a_p$.  

 Given the excellent correlation between the neutron skin and the $L_{sym}$ parameter, we can expect that adding an extra filter on the reproduction of the skin will allow to considerably reduce the uncertainty interval on $L_{sym}$
in a fully model-independent way. This study is not yet possible given the huge error bars of experimental data, 
but will hopefully be possible in a near future. 
\\

\section{Summary and Outlook}

In this work we have developed a meta-modeling analysis of the correlations of the empirical parameters among themselves and with nuclear observables such as masses, radii and neutron skins.

We used the Extended Thomas Fermi approximation at the second order in $\hbar$ and parametrised density profiles, to construct a fully analytical mass model for finite nuclei, based on a %empirical functional 
 {meta-modeling}
for homogeneous nuclear matter. 
The coefficients of this functional are directly related to the empirical parameters and can be independently varied, thus avoiding any artificial correlation induced by the chosen functional form.
In finite nuclei, a single isoscalar extra parameter is required  to reasonably reproduce the
experimental measurements of nuclear masses all along the nuclear chart.  
Our results show that no physical correlations exist among the different empirical parameters as far as the reproduction of binding energy is concerned, and thus suggest that the correlations shown in the literature might arise from the specific functional form assumed for the energy density, in particular for Skyrme functionals.

Charge radii exhibit interesting correlations both with EoS parameters ($K_{sat}$ and $n_{sat}$), and with the properties of the nuclear surface, which are less universal and might depend on the details of the variational theory.
It will be interesting to try to disentangle these two aspects with dedicated calculations in the future.  

We also showed that it is possible to reasonably reproduce the present measurements of neutron skins in nuclei even without the contribution from the differences in surface diffuseness between protons and neutrons.
In agreement with some previous studies, we find that the neutron skin depends only on the slope of the symmetry energy $L_{sym}$ and thus represent an extremely important observable to constrain the nuclear equation of state for astrophysical applications.  This result stresses the importance of precise measurements of this key quantity in the next future\cite{prex,Tarbert}.
  \\

\section*{Acknowledgments}
This work was partially supported by the NewCompStar COST action MP1304.
DC acknowledges the financial support from the CNRS and LPC.
Discussions with L.Colo are gratefully acknowledged.

\end{document}